\documentclass[english]{IEEEtran}
\usepackage[T1]{fontenc}
\usepackage[latin9]{inputenc}
\usepackage{color}
\usepackage{float}
\usepackage{textcomp}
\usepackage{mathrsfs}
\usepackage{amsmath}
\usepackage{amsthm}
\usepackage{amssymb}
\usepackage{graphicx}
\usepackage{wasysym}

\makeatletter

\providecommand{\tabularnewline}{\\}
\floatstyle{ruled}
\newfloat{algorithm}{tbp}{loa}
\providecommand{\algorithmname}{Algorithm}
\floatname{algorithm}{\protect\algorithmname}

\theoremstyle{plain}
\newtheorem{thm}{\protect\theoremname}
\theoremstyle{definition}
\newtheorem{defn}[thm]{\protect\definitionname}
\theoremstyle{remark}
\newtheorem{rem}[thm]{\protect\remarkname}

\usepackage{algorithmic}

\usepackage{subfig}

\makeatother

\usepackage{babel}
\providecommand{\definitionname}{Definition}
\providecommand{\remarkname}{Remark}
\providecommand{\theoremname}{Theorem}

\begin{document}

\title{\textcolor{black}{Angular-Domain Selective Channel Tracking and Doppler
Compensation for High-Mobility mmWave Massive MIMO}}

\author{\textcolor{black}{\normalsize{}Guanying Liu$^{1}$, An Liu$^{1}$,
}\textit{\textcolor{black}{\normalsize{}Senior Member, IEEE}}\textcolor{black}{\normalsize{},
Rui Zhang$^{2}$,}\textit{\textcolor{black}{\normalsize{} Fellow,
IEEE}}\textcolor{black}{\normalsize{}, and Min-Jian Zhao$^{1}$, }\textit{\textcolor{black}{\normalsize{}Member,
IEEE}}\textcolor{black}{\normalsize{}\\$^{1}$College of Information
Science and Electronic Engineering, Zhejiang University\\$^{2}$Department
of Electrical and Computer Engineering, National University of Singapore}}
\maketitle
\begin{abstract}
\textcolor{black}{In this paper, we consider a mmWave massive multiple-input
multiple-output (MIMO) communication system with one static base station
(BS) serving a fast-moving user, both equipped with a very large array.
The transmitted signal arrives at the user through multiple paths,
each with a different angle-of-arrival (AoA) and hence Doppler frequency
offset (DFO), thus resulting in a fast time-varying multipath fading
MIMO channel. In order to mitigate the Doppler-induced channel aging
for reduced pilot overhead, we propose a new angular-domain selective
channel tracking and Doppler compensation scheme at the user side.
Specifically, we formulate the joint estimation of partial angular--domain
channel and DFO parameters as a dynamic compressive sensing (CS) problem.
Then we propose a Doppler-aware-dynamic variational Bayesian inference
(DD-VBI) algorithm to solve this problem efficiently. Finally, we
propose a practical DFO compensation scheme which selects the dominant
paths of the fast time-varying channel for DFO compensation and thereby
converts it into a slow time-varying effective channel. Compared with
the existing methods, the proposed scheme can enjoy the huge array
gain provided by the massive MIMO and also balance the tradeoff between
the CSI signaling overhead and spatial multiplexing gain. Simulation
results verify the advantages of the proposed scheme over various
baseline schemes.}
\end{abstract}


\section{\textcolor{black}{Introduction}}

\textcolor{black}{With the development of the Fifth Generation (5G)
wireless communication systems, high-mobility scenarios such as high-speed
rail and Vehicle-To-everything (V2x) communications have gained increasingly
more interest. Due to the high-speed relative motion between the transmitter
and receiver, the transmitted signal, propagating through multiple
different paths, arrives at the receiver with different Doppler frequency
offsets (DFOs), thus resulting in a fast time-varying multipath fading
channel. In this case, the link performance such as achievable data
rate will be degraded significantly due to the Doppler-induced channel
aging effect \cite{Souden2009Robust}. To overcome this challenge,
several methods have been proposed in previous works, as elaborated
below. }

\textbf{\textcolor{black}{Direct Channel Estimation/Prediction: }}\textcolor{black}{For
line-of-sight (LOS) channels with a single LOS path, it is relatively
easy to compensate the Doppler effect and resolve the channel aging
issue by estimating the DFO parameter of the LOS path. However, when
there are multiple different paths due to rich scattering, it is challenging
to compensate the Doppler effect because different paths with different
DFOs are mixed together in the received signal. Nevertheless, some
works have proposed to directly estimate the fast time-varying channels
in the time/frequency domain \cite{Berger2010Application,Wang2018Channel}.
Existing channel estimators can be classified into two types. The
first type approximates time-varying channels using a linearly time-varying
(LTV) channel model \cite{Mostofi2005ICI,Liu2015On}. For example,
a hybrid frequency/time-domain channel estimation algorithm is proposed
in \cite{Mostofi2005ICI} based on the LTV model and two methods are
introduced to mitigate the Doppler effect. However, this algorithm
introduces a processing delay of at least one Orthogonal Frequency
Division Multiplexing (OFDM) symbol. The second type of estimators
adopt the basis expansion model (BEM) \cite{Xin2014Study} to convert
the problem of estimating the channel impulse response (CIR) to that
of estimating the basis function weights \cite{Al2010A}. For example,
in \cite{Al2010A}, the channel estimation and Doppler mitigation
are jointly considered by exploiting the correlations in time and
frequency domains, and the basis function coefficients are estimated
via the linear minimum mean squared error (LMMSE) approach. However,
accurate knowledge on the maximum DFO is required to determine the
minimum order of basis function and the computational burden is also
heavy for multi-antenna systems \cite{8558718}. Moreover, the BEM
inevitably introduces approximation error to channel estimation due
to the imperfect model assumed.}

\textbf{\textcolor{black}{Orthogonal Time Frequency Space (OTFS) Modulation:
}}\textcolor{black}{OTFS modulation \cite{Monk2016OTFS} is an emerging
technique which is able to handle the fast time-varying channels.
This method modulates transmitted symbols in the delay-Doppler domain
instead of time/frequency domain as in traditional modulation techniques
such as OFDM.. The idea is to transform the time-varying channel into
a time-invariant channel in the delay-Doppler domain. Early works
on OTFS modulation focused on the single-input single-output (SISO)
systems \cite{Monk2016OTFS}. Later, OTFS is extended to multiple-input
multiple-output (MIMO) systems by transmitting consecutive impulses
with proper guard time between two adjacent ones to distinguish different
base station (BS) antennas \cite{Ramachandran2018MIMO}. However,
the channel estimation method in \cite{Ramachandran2018MIMO} cannot
be directly applied to massive MIMO system since a large number of
antennas are required to be distinguished by transmitting such impulses,
which will lead to large pilot overhead.}

\textbf{\textcolor{black}{Angular-Domain DFO Estimation and Compensation:}}\textcolor{black}{{}
Since the different DFOs of multiple paths are resulted from their
different angles of arrival (AoAs)/angles of departure (AoDs) at the
receiver, they can be separated in the angle domain via spatial processing.
As such, angular-domain DFO estimation and compensation is another
popular approach to address the Doppler-induced channel aging issue
\cite{Souden2009Robust,Bellili2014A}. For MIMO systems, prior work
\cite{Chizhik2004Slowing} first pointed out that channel time-variation
can be slowed down through beamforming with a large number of transmit/receive
antennas. Motivated by this, a small-scale uniform circular antenna-array
(UCA) is adopted in \cite{Zhang2012Multiple,Guo2013Multiple} to separate
multiple DFOs via array beamforming. However, the DFO compensation
methods in \cite{Zhang2012Multiple,Guo2013Multiple} only apply to
scenarios with very sparse channels due to the limited spatial resolution
of small-scale MIMO. Recently, some works have exploited high-spatial
resolution provided by massive MIMO to address the high-mobility induced
challenges \cite{Liu2014On,Chen2017Directivity,Guo2017High,Guo2018High,Guo2018Angle}.
For example, the authors of \cite{8558718} propose to separate the
DFOs in angular-domain by beamforming with a large-scale uniform linear
antenna-array (ULA) at the mobile user side. After estimating and
compensating the DFO in each angle, the resultant quasi time-invariant
channel can be estimated more efficiently. However, the signaling
overhead for the maximum likelihood (ML) based joint estimation of
the massive MIMO channel matrix and DFO parameters is extremely high.
Moreover, only a single data stream is transmitted from the BS to
the mobile user via all possible channel directions, and thus it cannot
enjoy the spatial multiplexing gain as well as the huge array signal-to-noise
ratio (SNR) gain provided by massive MIMO. }

\textcolor{black}{The above works have focused on channel estimation.
It is also possible to directly search for the best beamforming vectors
without explicit channel estimation. For example, in \cite{Wang2009Beam},
the authors propose an exhaustive search (ES) scheme, which examines
all beam pairs in the codebook and determines the best pair that maximizes
a given performance metric (e.g., beamforming gain). To reduce the
training overhead of the ES scheme, the hierarchical search (HS) scheme
proposed in \cite{Alkhateeb2017Channel} utilizes hierarchical codebooks
and has a favorable performance at low SNRs. However, these codebook-based
beam search methods suffer from the quantization error caused by the
codebook and the channel aging effect. }

\textcolor{black}{In this paper, we consider high-mobility mmWave
massive MIMO systems, where both the BS and users are equipped with
massive antenna arrays to facilitate Doppler compensation, improve
the spectrum and energy efficiency, as well as overcome the large
path loss at high frequency band. As such, combining the mmWave and
massive MIMO technologies has the potential to significantly improve
the capacity and reliability of high-mobility wireless communications.
However, it is very challenging to design an efficient channel estimation
scheme. For example, since both ends have massive MIMO, the dimension
of the channel matrix is huge and conventional downlink/uplink channel
estimation or codebook-based beam search will lead to large signaling
overhead. Although various pilot overhead reduction methods such as
those based on compressive sensing (CS) have been proposed for the
estimation of slow massive MIMO fading channels \cite{Gao2016Channel}\cite{Bajwa2010Compressed},
they do not consider the Doppler effect and thus cannot be applied
to the high mobility scenario. To overcome this challenge, we propose
a novel angular-domain selective channel tracking and Doppler compensation
scheme, which exploits the dynamic sparsity of the mmWave massive
MIMO channel as well as precoded training in both downlink and uplink
to significantly reduce the signaling overhead. The main new contributions
of our paper are given as follows.}
\begin{itemize}
\item \textbf{\textcolor{black}{Angular-Domain Selective Channel Tracking:
}}\textcolor{black}{We propose a selective channel tracking scheme
to only estimate partial angular--domain channel parameters at the
user side that are sufficient for Doppler effect compensation to significantly
reduce the pilot overhead. Moreover, we propose an efficient downlink
training vector design at the BS side to strike a balance between
}\textit{\textcolor{black}{exploitation}}\textcolor{black}{{} of most
promising channel directions for array (SNR) gain and }\textit{\textcolor{black}{exploration}}\textcolor{black}{{}
of unknown channel directions. Compared to the conventional random
training vector design, the proposed design can exploit the massive
MIMO array gain to further enhance the channel tracking performance.}
\item \textbf{\textcolor{black}{Angular-Domain Selective Doppler Compensation:
}}\textcolor{black}{We propose an angular-domain selective DFO compensation
scheme at the user side which selectively converts the dominant paths
of the fast time-varying channel into a slow time-varying effective
channel. Compared to the non-selective DFO compensation scheme in
\cite{Guo2018High}, the proposed scheme can enjoy the huge array
(SNR) gain provided by the massive MIMO and also balance the tradeoff
between the CSI signaling overhead reduction and spatial multiplexing
gain maximization.}
\item \textbf{\textcolor{black}{Channel Tracking Algorithm based on Dynamic
VBI: }}\textcolor{black}{To further reduce the pilot overhead, the
proposed partial channel tracking design is formulated as a dynamic
CS problem with unknown DFO parameters in the measurement matrix.
Then, we adopt a three-layer hierarchical Markov model to capture
the dynamic sparsity of the partial angular--domain channel. The
existing methods, such as Variational Bayesian inference (VBI) \cite{Tzikas2008The}
and Sparse Bayesian Learning (SBL) \cite{Dai2018FDD}, cannot be directly
applied to this three-layer hierarchical prior. To address this challenge,
we propose a Doppler-aware-dynamic Variational Bayesian inference
(DD-VBI) algorithm, which combines the VBI and message-passing approaches
to achieve superior channel tracking performance.}
\end{itemize}
\textcolor{black}{}

\textcolor{black}{The rest of this paper is organized as follows.
In Section \ref{sec:System-Model}, we describe the system model and
frame structure. In Section \ref{sec:Angular-Domain-Selective-Channel},
we give a brief introduction of the proposed angular-domain selective
channel tracking and Doppler compensation scheme. In Sections \ref{sec:Downlink-Partial-Angular-Domain}
and \ref{sec:Doppler-Aware-Dynamic-VBI-algori}, we present the three-layer
hierarchical Markov model for partial angular channel vector and the
proposed DD-VBI algorithm. The simulation results and conclusions
are given in Sections \ref{sec:Simulation-Results} and \ref{sec:Conclusion}.}

\textit{\textcolor{black}{Notations:}}\textcolor{black}{{} For a set
of scalars $\left\{ x_{1},...,x_{N}\right\} $ and an index set $\mathcal{S}\subseteq\left\{ 1,...,N\right\} $,
we use $\left[x_{n}\right]_{n\in\mathcal{S}}$ to denote a column
vector consisting of the elements of $\left\{ x_{1},...,x_{N}\right\} $
indexed by the set $\mathcal{S}$. Similarly, for a set of column
vectors $\left\{ \mathbf{x}_{1},...,\mathbf{x}_{N}\right\} $ with
$\mathbf{x}_{n}\in\mathbb{C}^{M}$, $\left[\mathbf{x}_{n}\right]_{n\in\mathcal{S}}$
denotes a column vector consisting of the elements of $\left\{ \mathbf{x}_{1},...,\mathbf{x}_{N}\right\} $
indexed by the set $\mathcal{S}$. We use $\mathbf{X}\left(:,j\right)$
to denote the $j$-th column of a matrix $\mathbf{X}$. The key notations
are summarized in Table \ref{tab:Notations-1}.}

\textcolor{black}{}
\begin{table}
\begin{centering}
\textcolor{black}{}%
\begin{tabular}{|c|c|}
\hline 
\textcolor{black}{Notations} & \textcolor{black}{Meaning}\tabularnewline
\hline 
\hline 
\textcolor{black}{$N_{p}$} & \textcolor{black}{Number of downlink training vectors}\tabularnewline
\hline 
\textcolor{black}{$\mathbf{v}_{t}$} & \textcolor{black}{Downlink training vector}\tabularnewline
\hline 
\textcolor{black}{$M(N)$} & \textcolor{black}{Number of antennas at the BS (user)}\tabularnewline
\hline 
\textcolor{black}{{} $L_{t}$} & \textcolor{black}{Number of propagation paths}\tabularnewline
\hline 
\textcolor{black}{$\alpha_{t,q}$} & \textcolor{black}{The complex path gain of the $q$-th path}\tabularnewline
\hline 
\textcolor{black}{$f_{d,t}$} & \textcolor{black}{The maximum DFO}\tabularnewline
\hline 
\textcolor{black}{$\xi_{t,q}(\vartheta_{t,q})$} & \textcolor{black}{The AoD (AoA) of the $q$-th path}\tabularnewline
\hline 
\textcolor{black}{$\eta_{t}$} & \textcolor{black}{Rotation angle of user's antenna array}\tabularnewline
\hline 
\textcolor{black}{$\theta_{T,m}(\theta_{R,m})$} & \textcolor{black}{$m$-th AoD grid (AoA grid)}\tabularnewline
\hline 
\textcolor{black}{$\boldsymbol{\beta}_{T,t}(\boldsymbol{\beta}_{R,t})$} & \textcolor{black}{The AoD/AoA off-grid vector}\tabularnewline
\hline 
\textcolor{black}{$N_{b}$} & \textcolor{black}{Number of RF chains at the user}\tabularnewline
\hline 
\textcolor{black}{$\tilde{N}$} & \textcolor{black}{Number of AoA grid}\tabularnewline
\hline 
\end{tabular}
\par\end{centering}
\textcolor{black}{\caption{\textcolor{blue}{\label{tab:Notations-1}}\textcolor{black}{The key
notations used in the paper.}}
}
\end{table}

\section{\textcolor{black}{System Model\label{sec:System-Model}}}

\subsection{\textcolor{black}{System Architecture and Frame Structure}}

\textcolor{black}{Consider a time-division duplexing (TDD) mmWave
massive MIMO system with one static BS serving a fast-moving user}\footnote{\textcolor{black}{For clarity, we focus on a single user system. However,
the proposed selective channel tracking and Doppler compensation scheme
can be readily extended to multi-user systems.}}\textcolor{black}{. The BS is equipped with $M\gg1$ antennas. The
user is equipped with $N\gg1$ antennas. The time is divided into
frames, with each frame containing a downlink subframe and an uplink
subframe, as illustrated in Fig. 1. Each subframe contains a large
number of symbol durations.}

\textcolor{black}{}
\begin{figure}[htbp]
\begin{centering}
\textsf{\textcolor{black}{\includegraphics[scale=0.8]{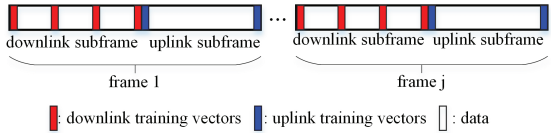}}}
\par\end{centering}
\textcolor{black}{\caption{Illustration of frame strcture.}
}
\end{figure}

\textcolor{black}{In the $t$-th downlink subframe, there are $N_{p}$
uniformly distributed training vectors, which are set to be the same
vector denoted as $\mathbf{v}_{t}$. For convenience, we use $\mathcal{N}_{p}$
to denote the symbol index set for the $N_{p}$ training vectors.
Note that inserting $N_{p}$ identical training vectors uniformly
in the downlink subframe facilitates the estimation of AoAs and Doppler
parameters at the user, as will be explained later. Based on the estimated
AoAs and Doppler parameters, the user applies a Doppler compensation
matrix to mitigate the Doppler effect and essentially converts the
fast time-varying channel into a slow time-varying effective channel.
In the uplink subframe, there are two sets of $N_{p}^{u}$ training
vectors at the beginning and end of the uplink subframe, respectively.
The two sets of uplink training vectors are used to estimate the slow
time-varying effective channel after Doppler compensation. Specifically,
the uplink transmission (e.g., beamforming and power allocation) in
the $t$-th uplink subframe is optimized based on the slow time-varying
effective channel estimated at the $t$-th uplink subframe. On the
other hand, by making use of the channel reciprocity, the downlink
transmission in the $t$-th downlink subframe is optimized based on
the slow time-varying effective channel estimated at the end of the
$\left(t-1\right)$-th uplink subframe. Since the effective channel
after Doppler compensation changes slowly compared to the subframe
duration, such a design can effectively overcome the channel aging
issue caused by the Doppler effect.}

\subsection{\textcolor{black}{Doppler Multipath Channel Model}}

\textcolor{black}{For clarity, we focus on the case when both the
BS and mobile user are equipped with a half-wavelength space ULA and
the channel is flat fading. To incorporate the DFO with conventional
mmWave channels, the downlink channel model for the antenna pair $\left\{ n_{t},n_{r}\right\} $
is given by \cite{Bajwa2010Compressed}}

\textcolor{black}{
\begin{equation}
{\color{blue}{\color{black}h_{n_{r},n_{t}t,i}=\sum_{q=1}^{L_{t}}\alpha_{t,q}e^{j\left[2\pi f_{d,t}icos(\vartheta_{t,q}+\eta_{t})+\psi_{n_{t}}(\xi_{t,q})+\psi_{n_{r}}(\vartheta_{t,q})\right]},}}\label{eq:jake's channel model}
\end{equation}
where $t$ stands for the frame index, $i$ stands for the symbol
index, $L_{t}$ is the total number of propagation paths, $\alpha_{t,q}$
is the random complex path gain associated with the $q$-th propagation
path, $f_{d,t}$ is the maximum DFO of the $t$-th frame, $\xi_{t,q}$
and $\vartheta_{t,q}$ are the AoD and AoA of the $q$-th path, respectively,
and $\eta_{t}$ is rotation angle of the user\textquoteright s antenna
array with respect to the moving direction in the t-th frame. Here,
$\psi_{n_{t}}(\xi_{t,q})$ and $\psi_{n_{r}}(\vartheta_{t,q})$ represent
the phase shifts induced at the $n_{t}$-th transmit antenna and the
$n_{r}$-th receive antenna, respectively, which depend on the antenna
structure, position, and the direction of the path. Note that in (\ref{eq:jake's channel model}),
we have implicitly assumed that the channel parameters $L_{t},\alpha_{t,q},\xi_{t,q},\vartheta_{t,q},f_{d,t},\eta_{t}$
are fixed within each frame but may change over different frames,
which is usually true even for high-speed users \cite{Guo2018High}.
However, the channel $h_{n_{r},n_{t},t,i}$ itself may change over
different symbols at a much faster timescale due to the fast changing
phase term $2\pi f_{d,t}icos(\vartheta_{t,q}+\eta_{t})$ caused by
the Doppler effect.}

\subsection{\textcolor{black}{Angular Domain Channel Representation}}

\textcolor{black}{To obtain the angular domain channel representation,
we introduce a uniform grid of $\tilde{M}$ AoDs and $\tilde{N}$
AoAs over $\left[0,2\pi\right)$}

\textcolor{black}{
\begin{multline*}
\{\theta_{T,m}:\\
sin(\theta_{T,m})=\frac{2}{\tilde{M}}\left(m-\left\lfloor \frac{\tilde{M}-1}{2}\right\rfloor \right),m=0,\ldots,\tilde{M}-1\},
\end{multline*}
}

\textcolor{black}{
\begin{multline*}
\{\theta_{R,n}:\\
sin(\theta_{R,n})=\frac{2}{\tilde{N}}\left(n-\left\lfloor \frac{\tilde{N}-1}{2}\right\rfloor \right),n=0,\ldots,\tilde{N}-1\}.
\end{multline*}
 }

\textcolor{black}{In practice, the true AoDs/AOAs usually do not lie
exactly on the grid points. In this case, there will be mismatches
between the true AoDs/AOAs and the nearest grid point. To overcome
this issue, we introduce an off-grid basis for the angular domain
channel representation, as in \cite{Dai2018FDD}. Specifically, let
$\theta_{T,m_{t,q}}$ and $\theta_{R,n_{t,q}}$ denote the nearest
grid point to $\xi_{t,q}$ and $\vartheta_{t,q}$, respectively. We
introduce the AoD off-grid vector $\boldsymbol{\beta}_{T,t}=\left[\beta_{T,t,1},\beta_{T,t,2},...,\beta_{T,t,\tilde{M}}\right]^{T}$
such that}

\textcolor{black}{
\[
\beta_{T,t,m}=\begin{cases}
\xi_{t,q}-\theta_{T,m_{t,q}}, & m=m_{t,q},q=1,2,...,L_{t}\\
0, & \text{otherwise}
\end{cases}.
\]
Similarly, let $\boldsymbol{\beta}_{R,t}=\left[\beta_{R,t,1},\beta_{R,t,2},...,\beta_{R,t,\tilde{N}}\right]^{T}$
denote the AoA off-grid vector, such that}

\textcolor{black}{
\[
\beta_{R,t,n}=\begin{cases}
\vartheta_{t,q}-\theta_{R,n_{t,q}}, & n=n_{t,q},q=1,2,...,L_{t}\\
0, & \text{otherwise}
\end{cases}.
\]
}

\textcolor{black}{For half-wavelength space ULAs, the array response
vectors at the BS and user side are given by $\boldsymbol{a}_{T}(\theta)=\frac{1}{\sqrt{M}}\left[1,e^{-j\pi sin(\theta)},e^{-j2\pi sin(\theta)},\ldots,e^{-j(M-1)\pi sin(\theta)}\right]^{T}$
and $\boldsymbol{a}_{R}(\theta)=\frac{1}{\sqrt{N}}\left[1,e^{-j\pi sin(\theta)},e^{-j2\pi sin(\theta)},\ldots,e^{-j(N-1)\pi sin(\theta)}\right]^{T}.$
For convenience, define two matrices ${\color{blue}\boldsymbol{A}_{R,i}\,(\boldsymbol{\beta}_{R,t},f_{d,t},\eta_{t})\:=\:[\tilde{\boldsymbol{a}}_{R,i}\,(\beta_{R,t,1},f_{d,t},\eta_{t}),\,\ldots\,,}$}

\noindent \textcolor{black}{${\color{blue}\tilde{\boldsymbol{a}}_{R,i}\,(\beta_{R,t,\tilde{N}},f_{d,t},\eta_{t})\,]\text{\ensuremath{\in}}\mathbb{C}^{N\text{\texttimes}\tilde{N}}}$
and $\boldsymbol{A}_{T}(\boldsymbol{\beta}_{T,t})=[\boldsymbol{a}_{T}(\theta_{T,1}+\beta_{T,t,1}),...,\boldsymbol{a}_{T}(\theta_{T,\tilde{M}}+\beta_{T,t,\tilde{M}})]\in\mathbb{C}^{M\times\tilde{M}}$,
where ${\color{blue}\tilde{\boldsymbol{a}\,}_{R,i}(\,\beta_{R,t,n},\,f_{d,t},\eta_{t})\,\,=}$}

\noindent \textcolor{black}{${\color{blue}\boldsymbol{a}_{R}\,(\theta_{R,n}+\,\beta_{R,t,n})\:\times\:e^{j2\pi f_{d,t}icos(\theta_{R,n}+\beta_{R,t,n}+\eta_{t})}}$.
Furthermore, define $\tilde{\boldsymbol{X}}_{t}\in\mathbb{C}^{\tilde{N}\times\tilde{M}}$
as the angular domain channel matrix with the $(n,m)$-th element
given by
\[
\tilde{x}_{t,n,m}=\begin{cases}
\alpha_{t,q}, & (n,m)=(n_{t,q},m_{t,q}),q=1,2,...,L_{t}\\
0, & \text{otherwise}
\end{cases}.
\]
Then, for given AoA off-grid, DFO parameter , rotation angle pair
$\boldsymbol{\varphi}_{t}=\left\{ \boldsymbol{\beta}_{R,t},f_{d,t},\eta_{t}\right\} $,
and AoD off-grid vector $\boldsymbol{\beta}_{T,t}$, $\boldsymbol{H}_{t,i}$
can be expressed in a compact form as}

\textcolor{black}{
\begin{equation}
\boldsymbol{H}_{t,i}\left(\boldsymbol{\varphi}_{t},\boldsymbol{\beta}_{T,t}\right)=\boldsymbol{A}_{R,i}(\boldsymbol{\varphi}_{t})\tilde{\boldsymbol{X}_{t}}\boldsymbol{A}_{T}^{H}(\boldsymbol{\beta}_{T,t}),\label{eq:angular domain}
\end{equation}
where t $e^{j\psi_{n_{t}}(\xi_{t,q})}$ and $e^{j\psi_{n_{r}}(\vartheta_{t,q})}$
in (\ref{eq:jake's channel model}) are implicitly contained in the
array response matrices $\boldsymbol{A}_{R,i}(\boldsymbol{\varphi}_{t})$
and $\boldsymbol{A}_{T}^{H}(\boldsymbol{\beta}_{T,t})$.}

\textcolor{black}{Note that we can also define the angular domain
representation for more general 2-dimensional (2D) antenna arrays.
In this case, the array response vector $\boldsymbol{a}_{T}(\theta,\phi)$
(or $\boldsymbol{a}_{R}(\theta,\phi)$) can be expressed as a function
of the azimuth angle $\theta$ and elevation angle $\phi$. Please
refer to \cite{Dietrich2000Adaptive} for the details.}

\section{\textcolor{black}{Angular-Domain Selective Channel Tracking and Doppler
Compensation\label{sec:Angular-Domain-Selective-Channel}}}

\textcolor{black}{In this section, we propose an efficient angular-domain
selective channel tracking and Doppler compensation scheme at the
user side. The proposed scheme can exploit both dynamic sparsity of
mmWave massive MIMO channel and high resolution of AoA at multi-antenna
mobile users to accurately estimate the downlink AoAs and maximum
DFO. Using these estimated parameters, a Doppler compensation matrix
is applied at the user to convert the fast time-varying channel into
a slow time-varying effective channel, based on which efficient downlink/uplink
transmissions can be achieved. The proposed scheme includes four key
components, namely the Angular-Domain Selective Channel Tracking,
Selective Doppler Compensation, Slow Time-Varying Effective Channel
Estimation, and Downlink Training Vector Design. Fig. 2 illustrates
a top-level diagram of the proposed scheme and the details of each
component are elaborated below. The frame index $t$ will be omitted
when there is no ambiguity.}

\textcolor{black}{}
\begin{figure}[htbp]
\begin{centering}
\textsf{\textcolor{black}{\includegraphics[scale=0.77]{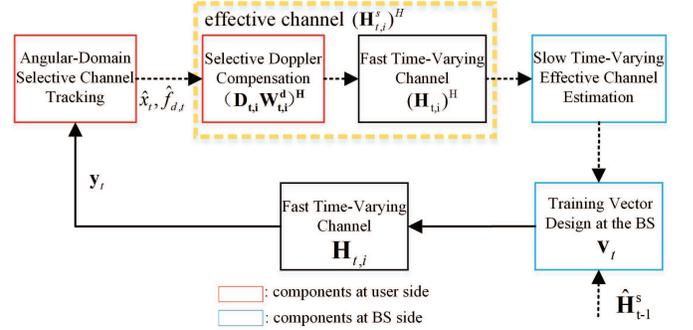}}}
\par\end{centering}
\textcolor{black}{\caption{A top-level diagram of the proposed scheme.}
}
\end{figure}

\subsection{\textcolor{black}{Outline of Angular-Domain Selective Channel Tracking
at the User}}

\textcolor{black}{This component is used to estimate the downlink
AoAs, rotation angle and maximum DFO based on the $N_{p}$ downlink
training vectors. Thanks to the high-spatial resolution provided by
the large array at the user side, the user can distinguish DFOs associated
with different AOAs from multiple active paths. However, since both
the BS and the user are equipped with the large array, the parameter
space can be very large if we attempt to estimate the full angular--domain
channel parameters (i.e., the full angular domain channel matrix $\tilde{\boldsymbol{X}}$,
rotation angle $\eta$ and maximum DFO $f_{d}$). Since the DFO only
occurs at the mobile user side, we propose to only estimate partial
angular--domain channel parameters that are just sufficient to obtain
AoAs, rotation angle and maximum DFO for Doppler compensation. }

\textcolor{black}{Specifically, the product of channel $\boldsymbol{H}_{i}$
and downlink training vector $\mathbf{v}$ can be expressed as:}

\textcolor{black}{
\begin{align}
\boldsymbol{H}_{i}\mathbf{v} & =\sum_{n=1}^{\tilde{N}}\sum_{m=1}^{\tilde{M}}\tilde{x}_{n,m}{\color{red}{\color{black}\tilde{\boldsymbol{a}}_{R,i}(\boldsymbol{\varphi})\boldsymbol{a}_{T}^{H}(\theta_{T,m}+\beta_{T,m})}}\mathbf{v},\nonumber \\
 & =\sum_{n=1}^{\tilde{N}}x_{n}{\color{black}{\color{red}{\color{black}\tilde{\boldsymbol{a}}_{R,i}(\boldsymbol{\varphi})}}={\color{red}{\color{black}\boldsymbol{A}_{R,i}(\boldsymbol{\varphi})}}}\boldsymbol{x},
\end{align}
where $\boldsymbol{x}=\left[x_{1},...,x_{\tilde{N}}\right]^{T}$ with
$x_{n}=\sum_{m=1}^{\tilde{M}}\tilde{x}_{n,m}{\color{red}{\color{black}\boldsymbol{a}_{T}^{H}(\theta_{T,m}+\beta_{T,m})}}\mathbf{v}$
are called partial angular channel coefficients since $\boldsymbol{x}$
only contain partial information about the full angular channel $\tilde{\boldsymbol{X}}$.
Specifically, a non-zero $\left|x_{n}\right|^{2}$ with value larger
than the noise floor indicates that there is an active path to the
$n$-th AoA direction at the user side. Therefore, we only need to
estimate $\tilde{N}$ partial channel parameters $\boldsymbol{x}$,
the AoA off-grid vector $\boldsymbol{\beta}_{R}$, rotation angle
$\eta$ and maximum DFO $f_{d}$, which are much less than the original
$\tilde{N}\tilde{M}$ full channel parameters, the off-grid vector
, rotation angle and maximum DFO. Note that if the $N_{p}$ training
vectors are different, there will be $\tilde{N}N_{p}$ partial angular
channel coefficients, leading to a larger parameter space to be estimated.
Moreover, with uniformly distributed training vectors, the phase rotation
due to the Doppler term $e^{j2\pi f_{d}icos(\theta_{R,n}+\beta_{R,t,n}+\eta_{t})}$
is larger compared to the case when the $N_{p}$ training vectors
are squeezed in the beginning of the downlink subframe, leading to
a better estimation performance for the Doppler parameter $f_{d}$.
Therefore, such a selective channel tracking design based on $N_{p}$
uniformly distributed and identical training vectors can significantly
reduce the number of downlink training vectors $N_{p}$ required to
achieve accurate estimation of AoAs, rotation angle and Doppler parameters.}

\textcolor{black}{The received baseband pilot signal is given by}

\textcolor{black}{
\begin{equation}
\boldsymbol{y}_{i}=\boldsymbol{H}_{i}\mathbf{v}+\boldsymbol{n}_{i},\forall i\in\mathcal{N}_{p}\label{eq:rec}
\end{equation}
where $\mathbf{v}\in\mathbb{C}^{M}$ is the training vector for downlink
channel tracking, and $\mathbf{n}_{i}$ is the additive white Gaussian
noise (AWGN) with each element having zero mean and variance $\sigma^{2}$,
respectively. The exact choice of $\mathbf{v}$ is postponed to Section
\ref{subsec:Downlink-Training-Vector}.}

\textcolor{black}{The aggregate received pilot signal (channel measurements)
of all the $N_{p}$ downlink pilot symbols (training vectors) in the
$t$-th frame can be expressed in a compact form as}

\textcolor{black}{
\begin{equation}
\boldsymbol{y}=\left[\boldsymbol{H}_{i}\mathbf{v}+\boldsymbol{n}_{i}\right]_{i\in\mathcal{N}_{p}}.\label{eq:rec_vector}
\end{equation}
}

\textcolor{black}{Based on the received downlink training vectors,
the user obtains the estimated partial channel parameters $\hat{\boldsymbol{x}}$,
$\hat{\boldsymbol{\beta}}_{R}$, $\hat{\eta}$ and $\hat{f}_{d}$
using a selective channel tracking algorithm. The detailed problem
formulation and algorithm design for selective channel tracking scheme
are postponed to Section \ref{sec:Downlink-Partial-Angular-Domain}.}

\subsection{\textcolor{black}{Angular-Domain Selective Doppler Compensation at
the User}}

\textcolor{black}{This component is used to convert the fast time-varying
channel into a slow time-varying effective channel after obtaining
the estimated partial channel parameters $\hat{\boldsymbol{x}}$,
$\hat{\boldsymbol{\beta}}_{R}$, $\hat{\eta}_{t}$ and $\hat{f}_{d}$.
We first select a set of $N_{d}$ most significant AoA directions
with the largest energy, where the energy of the $n$-th AoA direction
${\color{black}\theta_{R,n}+\beta_{R,n}}$ is defined as $\left|\hat{x}_{n}\right|^{2}$.
The parameter $N_{d}\leq N$ is used to control the tradeoff between
the spatial multiplexing gain and the effective CSI signaling overhead
(i.e., the CSI signaling overhead required to obtain effective channel
$\boldsymbol{H}_{i}^{s}$ in (\ref{eq:effectH})). Let $\mathcal{N}_{d}\subseteq\left\{ 1,...,N\right\} $
denote the index set of the selected $N_{d}$ most significant AoA
directions. Then, in order to mitigate the Doppler effect and perform
per-AoA DFO compensation for each selected AoA direction, a DFO compensation
matrix $\mathbf{W}_{i}^{d}\mathbf{D}_{i}\in\mathbb{C}^{N\times N_{d}}$,
which also serves as beamforming matrix, is applied at the user side.
In this way, we can convert the fast time-varying channel $\boldsymbol{H}_{i}$
into a slow time-varying effective channel $\boldsymbol{H}_{i}^{s}\in\mathbb{C}^{N_{d}\text{\texttimes}M}$
as}

\textcolor{black}{
\begin{align}
\boldsymbol{H}_{i}^{s} & =\mathbf{D}_{i}^{H}(\mathbf{W}_{i}^{d})^{H}\boldsymbol{H}_{i},\label{eq:effectH}
\end{align}
where $\mathbf{W}_{i}^{d}=\left[\boldsymbol{a}_{R}(\theta_{R,n}+\beta_{R,n})\right]_{n\in\mathcal{N}_{d}}\in\mathbb{C}^{N\times N_{d}}$
and $\mathbf{D}_{i}=\text{Diag}\left(\left[e^{j2\pi\hat{f}_{d}icos(\theta_{R,n}+\beta_{R,n}{\color{blue}+\eta_{t}})}\right]_{n\in\mathcal{N}_{d}}\right)\in\mathbb{C}^{N_{d}\times N_{d}}$.}

\textcolor{black}{In the following, we explain why the Doppler effect
can be alleviated by applying the above DFO compensation matrix to
obtain an effective channel $\boldsymbol{H}_{i}^{s}$. For half-wavelength
space ULA, if there is no estimation error for the partial channel
parameters, we have
\begin{equation}
\boldsymbol{H}_{i}^{s}=\sum_{m=1}^{\tilde{M}}\left[\tilde{x}_{n,m}\right]_{n\in\mathcal{N}_{d}}\boldsymbol{a}_{T}^{H}(\theta_{T,m}+\beta_{T,m})+O\left(\frac{1}{\sqrt{N}}\right),\label{eq:Hsi}
\end{equation}
as $N\rightarrow\infty$ \cite{Caire_TIT13_JSDM}. From (\ref{eq:Hsi}),
$\boldsymbol{H}_{i}^{s}$ is constant within a frame if we ignore
the small order term $O\left(\frac{1}{\sqrt{N}}\right)$, i.e., the
Doppler effect can be completely eliminated for sufficiently large
$N$. This observation is also consistent with the results in \cite{Guo2018High}.}

\subsection{\textcolor{black}{Slow Time-Varying Effective Channel Estimation
at the BS}}

\textcolor{black}{The user can simply transmit $N_{d}$ orthogonal
pilots in the uplink training stage. Then the conventional Least Squares
(LS) based channel estimation method can be used at the BS to obtain
the estimated slow time-varying effective channel $\hat{\boldsymbol{H}}_{i}^{s}$.
Based on $\hat{\boldsymbol{H}}_{i}^{s}$, the BS can optimize the
precoder for both uplink and downlink transmissions. Note that the
optimization of MIMO precoder is a standard problem and there are
many existing solutions with different performance and complexity
tradeoff. Then, the optimized uplink precoder is fed back to the user
for uplink transmission. Since $N_{d}$ can be much less than $N$,
the feedback overhead for the optimized uplink precoder is acceptable
for practice.}

\subsection{\textcolor{black}{Training Vector Design at the BS\label{subsec:Downlink-Training-Vector}}}

\textcolor{black}{The training vector $\mathbf{v}_{t}$ at the BS
is designed according to the slow time-varying effective channel $\hat{\boldsymbol{H}}_{t-1}^{s}$
estimated at the end of the $(t-1)$-th uplink subframe. The basic
idea for training vector design is to strike a balance between }\textit{\textcolor{black}{exploitation}}\textcolor{black}{{}
of known channel directions (i.e., transmitting training signal over
the most promising channel directions with large effective channel
energy to achieve beamforming gain) and }\textit{\textcolor{black}{exploration}}\textcolor{black}{{}
of unknown channel directions (i.e., transmitting training signal
over other channel directions to detect unknown channel directions).
Since the effective channel $\boldsymbol{H}_{i}^{s}$ changes slowly,
the effective channel $\hat{\boldsymbol{H}}_{t-1}^{s}$ estimated
at the end of the $(t-1)$-th uplink subframe is expected to provide
valuable information for the most promising channel directions. On
the other hand, the information about the most promising channel directions
extracted from $\hat{\boldsymbol{H}}_{t-1}^{s}$ may not be perfect
due to the estimation error and CSI delay. In addition, some new direction
may arise in the next frame. Therefore, the other channel directions
should also be incorporated into the training vector to facilitate
the detection of unknown channel directions. }

\textcolor{black}{Specifically, we first project the estimated effective
channel $\hat{\boldsymbol{H}}_{t-1}^{s}$ onto an orthogonal basis
$\mathbf{B}^{s}=\left[\mathbf{b}_{1}^{s},...,\mathbf{b}_{M}^{s}\right]\in\mathbb{C^{\mathit{M\times M}}}$
to obtain the effective channel energy on each basis vector (quantized
channel direction) as $\lambda_{m}^{s}=\left\Vert \hat{\boldsymbol{H}}_{t-1}^{s}\mathbf{b}_{m}^{s}\right\Vert ^{2},\forall m$.
The basis matrix is chosen such that the projection vector $\boldsymbol{\lambda}^{s}=\left[\lambda_{1}^{s},...,\lambda_{M}^{s}\right]^{T}$
is as sparse as possible. For half-wavelength ULAs, we can simply
choose the basis matrix $\mathbf{B}^{s}$ as an $M\times M$ DFT matrix.
Then we find the index set of the most promising channel directions
as 
\begin{equation}
\mathcal{M}^{*}=\text{argmin}_{\mathcal{M}}\left|\mathcal{M}\right|,\text{ s.t. }\sum_{m\in\mathcal{M}}\lambda_{m}^{s}/\sum_{m=1}^{M}\lambda_{m}^{s}\geq\mu,\label{eq:M}
\end{equation}
where $\mu$ is a threshold which is chosen to be closed to 1. In
other words, the most promising channel directions contain $\mu$
fraction of the total effective channel energy. Let $N_{s}=\left|\mathcal{M}^{*}\right|$.
Finally, the training vector is given by }

\textcolor{black}{
\begin{alignat}{1}
\mathbf{v}_{t}= & \frac{\sqrt{\rho}}{\sqrt{N_{s}}}\sum_{m\in\mathcal{M}^{*}}e^{j\theta_{m}^{s}}\mathbf{b}_{m}^{s}\nonumber \\
 & +\frac{\sqrt{1-\rho}}{\sqrt{M-N_{s}}}\sum_{m\in\left\{ 1,...,M\right\} \backslash\mathcal{M}^{*}}e^{j\theta_{m}^{s}}\mathbf{b}_{m}^{s}.\label{eq:vtdesign}
\end{alignat}
where the first term in (\ref{eq:vtdesign}) exploites the information
about the most promising $N_{s}$ channel directions extracted from
$\hat{\boldsymbol{H}}_{t-1}^{s}$, $\rho$ is a system parameter which
determines the proportion of transmit power used to exploit the most
promising channel directions, the second term is used to detect the
other unknown channel directions, $\theta_{m}^{s}$ is randomly generated
from $\left[0,2\pi\right]$. }

\textcolor{black}{}
\begin{figure}[htbp]
\begin{centering}
\textsf{\textcolor{black}{\includegraphics[scale=0.57]{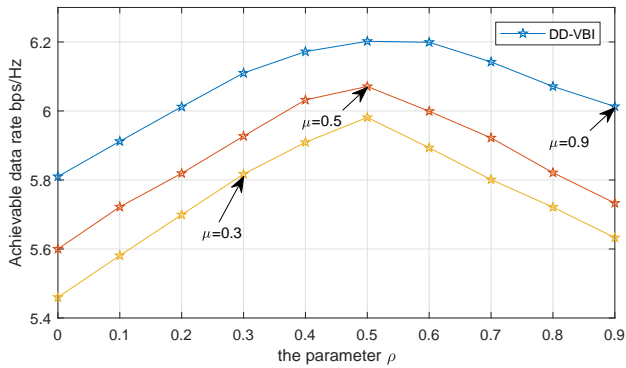}}}
\par\end{centering}
\textcolor{black}{\caption{\textcolor{blue}{\label{fig:Achievable-data-rate}}Achievable data
rate versus the parameter $\rho$ with different values of the parameter
$\mu$.}
}
\end{figure}

\textcolor{black}{In Fig. \ref{fig:Achievable-data-rate}, we illustrate
how the achievable data rate is affected by changing the parameters
$\rho$ and $\mu$ to achieve different tradeoffs between the exploration
and exploitation. It can be seen that setting $\mu=0.9$ and $\rho=0.5$
can strike a good balance between }\textit{\textcolor{black}{exploitation}}\textcolor{black}{{}
and }\textit{\textcolor{black}{exploration}}\textcolor{black}{.}

\section{\textcolor{black}{Problem Formulation for Angular-Domain Selective
Channel Tracking\label{sec:Downlink-Partial-Angular-Domain}}}

\subsection{\textcolor{black}{Three-layer Hierarchical Markov Model for Partial
Angular Channel Vector\label{subsec:Three-layer-Hierarchical-Markov}}}

\begin{figure}[htbp]
\begin{centering}
\textsf{\textcolor{black}{\includegraphics[scale=0.74]{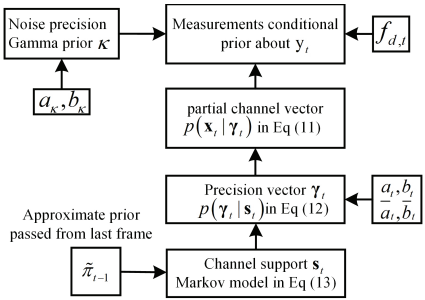}}}
\par\end{centering}
\textcolor{black}{\caption{\label{fig:Three-layer-hierarchical-Markov}Three-layer hierarchical
Markov model for partial angular channel coefficients.}
}
\end{figure}

\textcolor{black}{The dynamic sparsity of the partial angular channel
coefficients $\boldsymbol{x}_{t}$ is captured using a three-layer
hierarchical Markov model, as illustrated in Fig. \ref{fig:Three-layer-hierarchical-Markov}.
The first layer of random variable is the channel support vector $\boldsymbol{s}_{t}\in\{0,1\}^{\tilde{N}}$,
whose $n$-th element, denoted by $s_{t,n}$, indicates whether the
channel coefficient $x_{t,n}$ is active ($s_{t,n}=1$) or not ($s_{t,n}=0$).
The second layer of random variable is the precision vector $\boldsymbol{\gamma}_{t}=[\gamma_{t,1},\cdots,\gamma_{t,\tilde{N}}]^{T}$,
where $\gamma_{t,n}$ represents the precision (inverse of the variance)
of $x_{t,n}$. The third layer of random variables are partial angular
channel coefficients $\boldsymbol{x}_{t}$. For convenience, denote
a time series of vectors $\left\{ \boldsymbol{x}_{\tau}\right\} _{\tau=1}^{t}$
as $\boldsymbol{x}_{1:t}$ (same for $\mathbf{\boldsymbol{\gamma}}_{1:t}$,
$\boldsymbol{s}_{1:t}$, $f_{d,1:t}$,${\color{black}\boldsymbol{\beta}}_{{\color{black}R,1:t}}$).
Then the three-layer hierarchical Markov prior distribution (joint
distribution of $\boldsymbol{x}_{1:t}$, $\mathbf{\boldsymbol{\gamma}}_{1:t}$
and $\boldsymbol{s}_{1:t}$ ) is given by}

\textcolor{black}{
\begin{equation}
p(\boldsymbol{x}_{1:t},\mathbf{\boldsymbol{\gamma}}_{1:t},\boldsymbol{s}_{1:t})=\prod_{\tau=1}^{t}p\left(\boldsymbol{s}_{\tau}|\boldsymbol{s}_{\tau-1}\right)p(\mathbf{\boldsymbol{\gamma}}_{\tau}|\boldsymbol{s}_{\tau})p(\boldsymbol{x}_{\tau}|\mathbf{\boldsymbol{\gamma}}_{\tau}),\label{eq:three layer}
\end{equation}
where $p\left(\boldsymbol{s}_{1}|\boldsymbol{s}_{0}\right)\triangleq p\left(\boldsymbol{s}_{1}\right)$,
the conditional probability $p\left(\boldsymbol{x}_{\tau}|\mathbf{\boldsymbol{\gamma}}_{\tau}\right)$
has a product form $p\left(\boldsymbol{x}_{\tau}|\mathbf{\boldsymbol{\gamma}}_{\tau}\right)=\prod_{n=1}^{\tilde{N}}p\left(x_{\tau,n}|\mathbf{\gamma}_{\tau,n}\right)$
and each is modeled as a Gaussian prior distribution}

\textcolor{black}{
\begin{equation}
p\left(x_{\tau,n}|\gamma_{\tau,n}\right)=CN\left(x_{\tau,n};0,\gamma_{\tau,n}^{-1}\right),\label{eq:xcondruo}
\end{equation}
}

\textcolor{black}{The conditional prior of precision vector $\boldsymbol{\gamma}_{\tau}$
is given by}

\textcolor{black}{
\begin{equation}
p\left(\mathbf{\boldsymbol{\gamma}}_{\tau}|\boldsymbol{s}_{\tau}\right)=\prod_{n=1}^{\tilde{N}}\Gamma\left(\gamma_{\tau,n};a_{\tau},b_{\tau}\right)^{s_{\tau,n}}\Gamma\left(\gamma_{\tau,n};\overline{a}_{\tau},\overline{b}_{\tau}\right)^{1-s_{\tau,n}},\label{eq:gamma}
\end{equation}
$\Gamma\left(\gamma;a_{\gamma},b_{\gamma}\right)$ is a Gamma hyperprior.
$a_{\tau},b_{\tau}$ are the shape and rate parameters of the channel
precision $\gamma_{\tau,n}$ conditioned on $s_{\tau,n}=1$ and they
should be chosen such that $\frac{a_{\tau}}{b_{\tau}}=E[\gamma_{\tau,n}]=\Theta(1)$,
since the variance $\gamma_{\tau,n}^{-1}$ of $x_{\tau,n}$ is $\Theta(1)$
when it is active ($s_{\tau,n}=1$). $\overline{a}_{\tau},\overline{b}_{\tau}$
are the shape and rate parameters , conditioned on the opposite event
(i.e., $s_{\tau,n}=0$). In this case, the shape and rate parameters
$\overline{a}_{\tau},\overline{b}_{\tau}$ of the precision $\gamma_{\tau,n}$
should be chosen such that $\frac{\overline{a}_{\tau}}{\overline{b}_{\tau}}=E[\gamma_{\tau,n}]\gg1$,
since the variance $\gamma_{\tau,n}^{-1}$ of $x_{\tau,n}$ is close
to zero when it is inactive.}

\textcolor{black}{Note that the exact channel distribution is usually
unknown in practice. In this case, it is reasonable to choose a prior
distribution such that the derived algorithm can promote sparsity
with low complexity and achieve robust performance to different channel
distributions. By controlling the parameters in the Gamma distribution
of $\gamma_{\tau,n}$, one can easily promote sparsity based on the
knowledge of channel support $\boldsymbol{s}_{\tau}$, as explained
above. Moreover, Since the Gamma distribution for $\gamma_{\tau,n}$
is the conjugate probability distribution of the Gaussian distribution
for $x_{\tau,n}$, the above hierarchical prior for $x_{\tau,n}$
and $\gamma_{\tau,n}$ facilitates low-complexity VBI algorithm design
with closed-form update equations \cite{Tzikas2008The}. Finally,
the VBI-type algorithm derived from the such a hierarchical prior
is well known to be insensitive to the true distribution of the sparse
signals \cite{Tzikas2008The,Ji2008Bayesian}. As a result, similar
hierarchical prior distribution has been widely adopted in sparse
Bayesian learning \cite{Wipf2003Bayesian}.}

\textcolor{black}{Due to the slowly changing propagation environment,
the channel supports often change slowly over time, which implies
that $s_{\tau,n}$ depends on $s_{\tau-1,n}$, e.g., if $s_{\tau-1,n}=1$,
then there is a higher probability that $s_{\tau,n}$ is also 1. Such
dynamic sparsity of support vectors can be naturally modeled as a
temporal Markov model with an initial prior distribution $p(\mathbf{s}_{1})$
and a transition probability:}

\textcolor{black}{
\begin{equation}
p(\mathbf{s}_{\tau}|\mathbf{s}_{\tau-1})=\prod_{n=1}^{\tilde{N}}p(s_{\tau,n}|s_{\tau-1,n}),\label{eq:Markov support}
\end{equation}
where the transition probability is given by $p(s_{\tau,n}=1|s_{\tau-1,n}=0)=\rho_{0,1}$,
and $p(s_{\tau,n}=0|s_{\tau-1,n}=1)=\rho_{1,0}$. The Markov parameters
$\left\{ \rho_{1,0},\rho_{0,1}\right\} $ characterize the degree
of temporal correlation of the channel support. Specifically, smaller
$\rho_{1,0}$ or $\rho_{0,1}$ lead to highly correlated supports
across time, which means the propagation environment between the user
and BS is changing slowly. Larger $\rho_{1,0}$ or $\rho_{0,1}$ can
allow support to change substantially across time, which means the
propagation environment is changing significantly. Moreover, the statistic
parameters $\left\{ \rho_{1,0},\rho_{0,1}\right\} $ could be automatically
learned based on the EM framework during the recovery process \cite{Ziniel2013Dynamic},
as detailed in Appendix \ref{subsec:Derivation-2}. The initial distribution
$p(s_{1,n}),\forall n$ is set to be the steady state distribution
of the Markov chain in (\ref{eq:Markov support}), i.e.,}

\textcolor{black}{
\[
\lambda\triangleq p(s_{1,n})=\frac{\rho_{0,1}}{\rho_{0,1}+\rho_{1,0}}.
\]
This ensures that all elements of $s_{\tau,n}$ have the same marginal
distribution $p(s_{\tau,n})=\lambda^{s_{\tau,n}}(1-\lambda)^{1-s_{\tau,n}}$.}

\textcolor{black}{In practice, the noise precision $\kappa_{\tau}=\sigma_{\tau}^{-2}$
is usually unknown and we model it as a Gamma hyperpiror $p\left(\kappa_{\tau}\right)=\Gamma\left(\kappa_{\tau};a_{\kappa,\tau},b_{\kappa,\tau}\right)$,
where we set $a_{\kappa,\tau},b_{\kappa,\tau}\rightarrow0$ as in
\cite{Dai2018FDD} so as to obtain a broad hyperprior.}

\textcolor{black}{}

\subsection{\textcolor{black}{Selective Channel Tracking Formulation}}

\textcolor{black}{Using the angular domain channel representation,
the receive signal at $t$-th frame $\boldsymbol{y}_{t}\in\mathbb{C}^{NN_{p}}$
can be rewritten as a CS model with an unknown AoA off-grid and DFO
parameter pair ${\color{blue}\boldsymbol{\varphi}_{t}=\left\{ \boldsymbol{\beta}_{R,t},\eta_{t},f_{d,t}\right\} }$
in the measurement matrix as}

\textcolor{black}{
\begin{equation}
\boldsymbol{y}_{t}=\boldsymbol{F}_{t}\boldsymbol{x}_{t}+\boldsymbol{n}_{t},\label{eq:compress formulation}
\end{equation}
where the measurement matrix is given by $\boldsymbol{F}_{t}=[\boldsymbol{F}_{t,1};...;\boldsymbol{F}_{t,N_{p}}]\text{\ensuremath{\in}}C^{NN_{p}\text{\texttimes}\tilde{N}}$,
$\boldsymbol{F}_{t,i}=\boldsymbol{A}_{R,i}({\color{red}{\color{black}\boldsymbol{\varphi}_{t}}})$,
$\boldsymbol{n}_{t}=[\boldsymbol{n}_{t,i}]_{i\in\mathcal{N}_{p}}$.}

\textcolor{black}{In each frame $t$, the user needs to estimate the
partial channel parameters $\boldsymbol{x}_{t}$ , the AoA off-grid
and DFO parameter pair ${\color{blue}\boldsymbol{\varphi}_{t}=\left\{ \boldsymbol{\beta}_{R,t},\eta_{t},f_{d,t}\right\} }$,
given the observations up to $t$ frame $\boldsymbol{y}_{1:t}$ in
model (\ref{eq:compress formulation}), the estimated AoA off-grid
and DFO parameter pairs $\hat{\boldsymbol{\varphi}}_{1:t-1}=\left\{ \hat{\boldsymbol{\beta}}_{R,1:t-1},\hat{\eta}_{1:t-1},\hat{f}_{d,1:t-1}\right\} $
up to $\left(t-1\right)$ frame. In particular, for given $\boldsymbol{\varphi}_{t}$,
we are interested in computing minimum mean-squared error (MMSE) estimates
of ${x_{t,n}}$, $\hat{x}_{t,n}={\color{red}{\color{black}\mathrm{E}\left[x_{t,n}|\boldsymbol{y}_{1:t};\hat{\boldsymbol{\varphi}}_{1:t-1},\boldsymbol{\varphi}_{t}\right]}}$,
where the expectation is over the marginal posterior:}

\textcolor{black}{
\begin{flalign}
{\color{red}} & {\color{black}{\color{red}{\color{black}p(x_{t,n}|\boldsymbol{y}_{1:t};\hat{\boldsymbol{\varphi}}_{1:t-1},\boldsymbol{\varphi}_{t})}}}\nonumber \\
{\color{red}{\color{black}\propto}} & {\color{black}{\color{red}{\color{black}\int_{-x_{t,n}}p(\boldsymbol{y}_{1:t},\boldsymbol{v}_{t};\hat{\boldsymbol{\varphi}}_{1:t-1},\boldsymbol{\varphi}_{t}),}}}\label{eq:posterior}
\end{flalign}
where $\boldsymbol{v}_{t}=\left\{ \boldsymbol{x}_{t},\boldsymbol{s}_{t},\boldsymbol{\gamma}_{t},\kappa_{t}\right\} $,
$-x_{t,n}$ denotes the vector collections integration over $\boldsymbol{v}_{t}$
except for the element $x_{t,n}$ and $\propto$ denotes equality
after scaling.}

\textcolor{black}{On the other hand, the optimal $\boldsymbol{\varphi}_{t}$
at the $t$-th frame is obtained by ML as follows \cite{Dai2018FDD}:}

\textcolor{black}{
\begin{alignat}{1}
\boldsymbol{\hat{\varphi}}_{t} & {\color{red}{\color{black}=\mathrm{arg}\max_{\boldsymbol{\varphi}_{t}}\ln p(\boldsymbol{y}_{1:t};\hat{\boldsymbol{\varphi}}_{1:t-1},\boldsymbol{\varphi}_{t})}}\nonumber \\
{\color{red}} & {\color{red}{\color{black}\mathrm{=arg}\max_{\boldsymbol{\varphi}_{t}}\ln\int_{\boldsymbol{v}_{t}}p(\boldsymbol{y}_{1:t},\boldsymbol{v}_{t};\hat{\boldsymbol{\varphi}}_{1:t-1},\boldsymbol{\varphi}_{t})d\boldsymbol{v}_{t}.}}\label{eq:maxlikelyhood}
\end{alignat}
}

\textcolor{black}{Once we obtain the ML estimate of $\boldsymbol{\hat{\varphi}}_{t}$,
and the associated conditional marginal posterior $p(x_{t,n}|\boldsymbol{y}_{1:t};\hat{\boldsymbol{\varphi}}_{1:t-1},\boldsymbol{\varphi}_{t})$,
we can obtain the MMSE estimates of $\left\{ x_{t,n}\right\} $.}

\textcolor{black}{One challenge in computing the MMSE estimate is
the calculation of the exact posterior in (\ref{eq:posterior}) whose
factor graph has loops. In the next subsection, we propose a Doppler-aware-dynamic-VBI
(DD-VBI) algorithm to approximately calculate the marginal posteriors
${p(x_{t,n}|\boldsymbol{y}_{1:t};\hat{\boldsymbol{\varphi}}_{1:t-1},\boldsymbol{\varphi}_{t})}$
by combining the message passing and VBI approaches, and use the in-exact
majorization-minimization (MM) method (which is a generalization of
the EM method) \cite{Dai2018FDD} to find an approximate solution
for (\ref{eq:maxlikelyhood}). The proposed DD-VBI algorithm is shown
in the simulations to achieve a good performance.}

\section{\textcolor{black}{Doppler-Aware-Dynamic-VBI algorithm \label{sec:Doppler-Aware-Dynamic-VBI-algori}}}

\subsection{\textcolor{black}{Decomposition and Approximation of Joint Probability
Distribution\label{subsec:Decomposition-Joint-Probability}}}

\textcolor{black}{This section is to decompose and approximate the
joint probability distribution in (\ref{eq:maxlikelyhood}) such that
the joint probability distribution at the $t$-th frame only involves
the probability density function (PDF) of the current hidden variables
$\boldsymbol{v}_{t}$, the current observation $\boldsymbol{y}_{t}$,
and the messages $\hat{p}(\boldsymbol{s}_{t}|\boldsymbol{y}_{1:t-1},\hat{\boldsymbol{\varphi}}_{1:t-1})$
passed from the previous frame, based on which a more efficient algorithm
can be designed.}

\textcolor{black}{The joint probability distribution in (\ref{eq:maxlikelyhood})
and (\ref{eq:posterior}) can be written as}

\textcolor{black}{
\begin{align*}
 & p(\boldsymbol{y}_{1:t},\boldsymbol{v}_{t};\hat{\boldsymbol{\varphi}}_{1:t-1},\boldsymbol{\varphi}_{t})\\
\propto & \sum_{\boldsymbol{s}_{t-1}}p(\mathbf{s}_{t-1}|\mathbf{y}_{1:t-1};\hat{\boldsymbol{\varphi}}_{1:t-1})p(\boldsymbol{s}_{t}|\boldsymbol{s}_{t-1})\\
 & p(\boldsymbol{y}_{t}|\boldsymbol{x}_{t},\kappa_{t};\boldsymbol{\varphi}_{t})p(\boldsymbol{x}_{t}|\mathbf{\boldsymbol{\gamma}}_{t})p(\mathbf{\boldsymbol{\gamma}}_{t}|\boldsymbol{s}_{t})p(\kappa_{t})\\
\approx & \sum_{\boldsymbol{s}_{t-1}}q(\boldsymbol{s}_{t-1}|\boldsymbol{y}_{1:t-1};\hat{\boldsymbol{\varphi}}_{1:t-1})p(\boldsymbol{s}_{t}|\boldsymbol{s}_{t-1})\\
 & p(\boldsymbol{y}_{t}|\boldsymbol{x}_{t},\kappa_{t};\boldsymbol{\varphi}_{t})p(\boldsymbol{x}_{t}|\mathbf{\boldsymbol{\gamma}}_{t})p(\mathbf{\boldsymbol{\gamma}}_{t}|\boldsymbol{s}_{t})p(\kappa_{t})\\
= & \hat{p}(\boldsymbol{s}_{t}|\boldsymbol{y}_{1:t-1};\hat{\boldsymbol{\varphi}}_{1:t-1})p(\boldsymbol{y}_{t}|\boldsymbol{x}_{t},\kappa_{t};\boldsymbol{\varphi}_{t})\\
 & p(\boldsymbol{x}_{t}|\mathbf{\boldsymbol{\gamma}}_{t})p(\mathbf{\boldsymbol{\gamma}}_{t}|\boldsymbol{s}_{t})p(\kappa_{t}),
\end{align*}
where $\hat{p}(\boldsymbol{s}_{t}|\boldsymbol{y}_{1:t-1};\hat{\boldsymbol{\varphi}}_{1:t-1})=\sum_{\mathbf{s}_{t-1}}q(\boldsymbol{s}_{t-1}|\boldsymbol{y}_{1:t-1};\hat{\boldsymbol{\varphi}}_{1:t-1})p(\boldsymbol{s}_{t}|\boldsymbol{s}_{t-1})$,
$q(\boldsymbol{s}_{t-1}|\boldsymbol{y}_{1:t-1};\hat{\boldsymbol{\varphi}}_{1:t-1})$
is a tractable approximation for the posterior $p(\mathbf{s}_{t-1}|\mathbf{y}_{1:t-1};\hat{\boldsymbol{\varphi}}_{1:t-1})$
and $p\left(\boldsymbol{y}_{t}|\boldsymbol{x}_{t},\kappa_{t};\boldsymbol{\varphi}_{t}\right)=\mathcal{CN}\left(\boldsymbol{y}_{t};\boldsymbol{F}_{t}\boldsymbol{x}_{t},\kappa_{t}^{-1}\boldsymbol{I}\right)$.
Both $q(\boldsymbol{s}_{t-1}|\boldsymbol{y}_{1:t-1};\hat{\boldsymbol{\varphi}}_{1:t-1})$
and $\hat{p}(\boldsymbol{s}_{t}|\boldsymbol{y}_{1:t-1};\hat{\boldsymbol{\varphi}}_{1:t-1})$
can be calculated based on the messages passed from the previous frame.
We will elaborate how to calculate $\hat{p}(\boldsymbol{s}_{t}|\boldsymbol{y}_{1:t-1};\hat{\boldsymbol{\varphi}}_{1:t-1})$
later in subsection \ref{subsec:Cauculating--passed}. When $t=1$,
$\hat{p}(\boldsymbol{s}_{t}|\boldsymbol{y}_{1:t-1};\hat{\boldsymbol{\varphi}}_{1:t-1})$
is reduced to $\hat{p}(\boldsymbol{s}_{t}|\boldsymbol{y}_{1};\hat{\boldsymbol{\varphi}}_{1})=p(\boldsymbol{s}_{1})$.}

\textcolor{black}{For simplicity, we define
\begin{alignat}{1}
 & \hat{p}(\boldsymbol{y}_{1:t},\boldsymbol{v}_{t};\hat{\boldsymbol{\varphi}}_{1:t-1},\boldsymbol{\varphi}_{t})\nonumber \\
= & \hat{p}(\boldsymbol{s}_{t}|\boldsymbol{y}_{1:t-1};\hat{\boldsymbol{\varphi}}_{1:t-1})p(\boldsymbol{y}_{t}|\boldsymbol{x}_{t},\kappa_{t};\boldsymbol{\varphi}_{t})\nonumber \\
 & p(\boldsymbol{x}_{t}|\mathbf{\boldsymbol{\gamma}}_{t})p(\mathbf{\boldsymbol{\gamma}}_{t}|\boldsymbol{s}_{t})p(\kappa_{t}).\label{eq:pyvfappro}
\end{alignat}
}

\textcolor{black}{In the rest of this section, we will omit $\hat{\boldsymbol{\varphi}}_{1:t-1}$
in the PDFs when there is no ambiguity.}

\subsection{\textcolor{black}{Outline of the Doppler-Aware-Dynamic-VBI algorithm
in Frame $t$}}

\begin{figure}[htbp]
\begin{centering}
\textsf{\textcolor{black}{\includegraphics[scale=0.77]{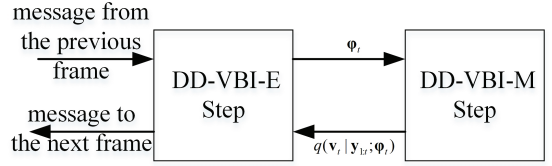}}}
\par\end{centering}
\textcolor{black}{\caption{I\label{fig:Interaction-between-the}interaction between the two modules
of the DD-VBI algorithm within a frame.}
}
\end{figure}

\textcolor{black}{The basic idea of the DD-VBI algorithm is that,
at every frame $t$, simultaneously approximates the marginal posterior
$\left\{ p(x_{t,n}|\boldsymbol{y}_{1:t};\boldsymbol{\varphi}_{t})\right\} $
and maximizes the log-likelihood $\ln p(\boldsymbol{y}_{1:t};\boldsymbol{\varphi}_{t})$
with respect to $\boldsymbol{\varphi}_{t}$, based on the noisy measurements
of the $t$-th frame and the messages $\hat{p}(\boldsymbol{s}_{t}|\boldsymbol{y}_{1:t-1})$
passed from the previous frame. In summary, for every frame, the DD-VBI
algorithm performs iterations between the following two major steps
until convergence, as shown in Fig. \ref{fig:Interaction-between-the}.}
\begin{itemize}
\item \textcolor{black}{DD-VBI-E Step: Given $\boldsymbol{\varphi}_{t}$
at $t$-th frame and messages $\hat{p}(\boldsymbol{s}_{t}|\boldsymbol{y}_{1:t-1};\hat{\boldsymbol{\varphi}}_{1:t-1})$
passed from the previous frame, calculate the approximate marginal
posterior of $p(\boldsymbol{v}_{t}|\boldsymbol{y}_{1:t};\boldsymbol{\varphi}_{t})$,
denoted as $q(\boldsymbol{v}_{t}|\boldsymbol{y}_{1:t};\boldsymbol{\varphi}_{t})$,
using the sparse VBI approach, as elaborated in subsection \ref{subsec:Doppler-Turbo-OAMP-E-Step}.}
\item \textcolor{black}{DD-VBI-M Step: Given $q(\boldsymbol{v}_{t}|\boldsymbol{y}_{1:t};\boldsymbol{\varphi}_{t})\approx p(\boldsymbol{v}_{t}|\boldsymbol{y}_{1:t};\boldsymbol{\varphi}_{t})$,
construct a surrogate function for the objective function $\ln p(\boldsymbol{y}_{1:t};\boldsymbol{\varphi}_{t})$,
then maximize the surrogate function with respect to $\boldsymbol{\varphi}_{t}$
as elaborated in subsection \ref{subsec:Doppler-Turbo-OAMP-M-Step}. }
\end{itemize}
\textcolor{black}{After convergence, the messages $\hat{p}(\boldsymbol{s}_{t+1}|\boldsymbol{y}_{1:t};\hat{\boldsymbol{\varphi}}_{1:t})$
are calculated based on $q(\boldsymbol{s}_{t}|\boldsymbol{y}_{1:t};\boldsymbol{\varphi}_{1:t})$
and passed to the next frame. In the following, we first elaborate
the M step, which is a variation of the in-exact MM method in \cite{Dai2018FDD}.
After that, we will elaborate how to approximately calculate the posterior
$p(\boldsymbol{v}_{t}|\boldsymbol{y}_{1:t};\boldsymbol{\varphi}_{t})\approx q(\boldsymbol{v}_{t}|\boldsymbol{y}_{1:t};\boldsymbol{\varphi}_{t})$
in the E step, which is required to construct the surrogate function
in the M step.}

\textcolor{black}{}

\subsection{\textcolor{black}{DD-VBI-M Step \label{subsec:Doppler-Turbo-OAMP-M-Step}}}

\textcolor{black}{It is difficult to directly maximize the log-likelihood
function $\ln p(\boldsymbol{y}_{1:t};\boldsymbol{\varphi}_{t})$,
because there is no closed-form expression due to the multi-dimensional
integration over $\boldsymbol{v}_{t}$ as in (\ref{eq:maxlikelyhood}).
To make the problem tractable, in the DD-VBI-M Step, we adopt an in-exact
MM method in \cite{A1977Maximum,Dai2018FDD}, which maximizes a surrogate
function of $\ln p(\boldsymbol{y}_{1:t};\boldsymbol{\varphi}_{t})$
with respect to $\boldsymbol{\varphi}_{t}$, to find an approximate
solution of (\ref{eq:maxlikelyhood}). Specifically, let $u(\boldsymbol{\varphi}_{t};\boldsymbol{\dot{\varphi}}_{t})$
be the surrogate function constructed at some fixed point $\boldsymbol{\dot{\varphi}}_{t}$,
which satisfies the following properties:}

\textcolor{black}{
\begin{gather}
u(\boldsymbol{\varphi}_{t};\boldsymbol{\dot{\varphi}}_{t})\leq\ln p(\boldsymbol{y}_{1:t};\boldsymbol{\varphi}_{t}),\nonumber \\
u(\boldsymbol{\dot{\varphi}}_{t};\boldsymbol{\dot{\varphi}}_{t})=\ln p(\boldsymbol{y}_{1:t};\boldsymbol{\dot{\varphi}}_{t}),\nonumber \\
\frac{\partial u(\boldsymbol{\varphi}_{t};\boldsymbol{\dot{\varphi}}_{t})}{\partial\boldsymbol{\varphi}_{t}}|_{\boldsymbol{\varphi}_{t}=\boldsymbol{\dot{\varphi}}_{t}}=\frac{\partial\ln p(\boldsymbol{y}_{1:t};\boldsymbol{\varphi}_{t})}{\partial\boldsymbol{\varphi}_{t}}|_{\boldsymbol{\varphi}_{t}=\boldsymbol{\dot{\varphi}}_{t}}.\label{eq:surr}
\end{gather}
Inspired by the EM algorithm \cite{A1977Maximum}, we use the following
surrogate function:}

\textcolor{black}{
\begin{equation}
u(\boldsymbol{\varphi}_{t};\boldsymbol{\dot{\varphi}}_{t})=\int q(\boldsymbol{v}_{t}|\boldsymbol{y}_{1:t};\boldsymbol{\varphi}_{t})\ln\frac{p(\boldsymbol{y}_{1:t},\boldsymbol{v}_{t};\boldsymbol{\varphi}_{t})}{q(\boldsymbol{v}_{t}|\boldsymbol{y}_{1:t};\boldsymbol{\varphi}_{t})}d\boldsymbol{v}_{t},\label{eq:surrogate function}
\end{equation}
where $q(\boldsymbol{v}_{t}|\boldsymbol{y}_{1:t};\boldsymbol{\varphi}_{t})$
is a tractable approximation of $p(\boldsymbol{v}_{t}|\boldsymbol{y}_{1:t};\boldsymbol{\varphi}_{t})$.
In subsection \ref{subsec:Decomposition-Joint-Probability}, we have
approximated the joint probability distribution in (\ref{eq:surrogate function})
using $\hat{p}(\boldsymbol{y}_{1:t},\boldsymbol{v}_{t};\boldsymbol{\varphi}_{t})$.
Therefore, the surrogate function can be approximated as}

\textcolor{black}{
\begin{equation}
\hat{u}(\boldsymbol{\varphi}_{t};\boldsymbol{\dot{\varphi}}_{t})=\int q(\boldsymbol{v}_{t}|\boldsymbol{y}_{1:t};\boldsymbol{\varphi}_{t})\ln\frac{\hat{p}(\boldsymbol{y}_{1:t},\boldsymbol{v}_{t};\boldsymbol{\varphi}_{t})}{q(\boldsymbol{v}_{t}|\boldsymbol{y}_{1:t};\boldsymbol{\varphi}_{t})}d\boldsymbol{v}_{t},\label{eq:surr_1}
\end{equation}
When there is no approximation error for the associated PDFs, i.e.,
$q(\boldsymbol{v}_{t}|\boldsymbol{y}_{1:t};\boldsymbol{\varphi}_{t})=p(\boldsymbol{v}_{t}|\boldsymbol{y}_{1:t};\boldsymbol{\varphi}_{t})$
and $\hat{p}(\boldsymbol{y}_{1:t},\boldsymbol{v}_{t};\boldsymbol{\varphi}_{t})=p(\boldsymbol{y}_{1:t},\boldsymbol{v}_{t};\boldsymbol{\varphi}_{t})$,
it can be verified that $\hat{u}(\boldsymbol{\varphi}_{t};\boldsymbol{\dot{\varphi}}_{t})$
satisfies the properties in (\ref{eq:surr}).}

\textcolor{black}{In the M step of the $j$-th iteration, we update
$\boldsymbol{\varphi}_{t}$ as
\begin{gather}
\boldsymbol{\varphi}_{t}^{j+1}=\mathrm{arg}\max_{\boldsymbol{\varphi}_{t}}\hat{u}(\boldsymbol{\varphi}_{t};\boldsymbol{\varphi}_{t}^{j}),\label{eq:max_surr}
\end{gather}
where $\left(\cdot\right)^{j}$ stands for the $j$-th iteration. }

\textcolor{black}{In our problem, $\hat{u}(\boldsymbol{\varphi}_{t};\boldsymbol{\dot{\varphi}}_{t})$
is a non-convex function and it is difficult to find its optimal solution.
Therefore, we use a simple gradient update as in \cite{Dai2018FDD},
i.e.}

\textcolor{black}{
\begin{align}
\boldsymbol{\varphi}_{t}^{j+1} & =\boldsymbol{\varphi}_{t}^{j}+\boldsymbol{\tau}^{j}\frac{\partial\hat{u}(\boldsymbol{\varphi}_{t};\boldsymbol{\dot{\varphi}}_{t}^{j})}{\partial\boldsymbol{\varphi}_{t}},\label{eq:ugrad}
\end{align}
where $\boldsymbol{\tau}^{j}$ is the step sizes determined by the
Armijo rule \cite{Dai2018FDD}.}

\textcolor{black}{The approximate posterior $q\left(\boldsymbol{v}_{t}|\boldsymbol{y}_{1:t};\boldsymbol{\varphi}_{t}\right)$
has a factorized form as
\begin{multline}
q\left(\boldsymbol{v}_{t}|\boldsymbol{y}_{1:t};\boldsymbol{\varphi}_{t}\right)\\
=q\left(\boldsymbol{x}_{t}|\boldsymbol{y}_{1:t};\boldsymbol{\varphi}_{t}\right)q\left(\boldsymbol{\gamma}_{t}|\boldsymbol{y}_{1:t};\boldsymbol{\varphi}_{t}\right)q\left(\boldsymbol{s}_{t}|\boldsymbol{y}_{1:t};\boldsymbol{\varphi}_{t}\right)\text{.}\label{eq:factorized}
\end{multline}
}

\textcolor{black}{Therefore, after the convergence of the DD-VBI,
we not only obtain an approximate stationary solution $\boldsymbol{\hat{\varphi}}_{t}$
of (\ref{eq:maxlikelyhood}), but also the associated (approximate)
marginal conditional posterior $q\left(\boldsymbol{x}_{t}|\boldsymbol{y}_{1:t};\boldsymbol{\varphi}_{1:t}\right)\approx p(\boldsymbol{x}_{t}|\boldsymbol{y}_{1:t},\boldsymbol{\varphi}_{t})$.}

\subsection{\textcolor{black}{DD-VBI-E Step\label{subsec:Doppler-Turbo-OAMP-E-Step}}}

\textcolor{black}{DD-VBI-E Step performs the sparse VBI to approximate
the conditional marginal posteriors $p(\boldsymbol{v}_{t}|\boldsymbol{y}_{1:t};\boldsymbol{\varphi}_{t})$
based on the following joint prior distribution: }

\textcolor{black}{
\begin{alignat}{1}
 & \hat{p}(\boldsymbol{y}_{1:t},\boldsymbol{v}_{t};\boldsymbol{\varphi}_{t})\nonumber \\
= & \hat{p}(\boldsymbol{s}_{t}|\boldsymbol{y}_{1:t-1})p(\boldsymbol{y}_{t}|\boldsymbol{x}_{t},\kappa_{t};\boldsymbol{\varphi}_{t})\nonumber \\
 & p(\boldsymbol{x}_{t}|\mathbf{\boldsymbol{\gamma}}_{t})p(\mathbf{\boldsymbol{\gamma}}_{t}|\boldsymbol{s}_{t})p(\kappa_{t}).\label{eq:pyvfappro-1}
\end{alignat}
}

\textcolor{black}{The corresponding approximate posterior distributions
$q\left(\boldsymbol{v}_{t}\right)$ obtained by the sparse VBI will
be given by (\ref{eq:poster_kappa})-(\ref{eq:poster_q}).}

\subsubsection{\textcolor{black}{Outline of Sparse VBI within a Frame}}

\textcolor{black}{For convenience, we use $\boldsymbol{v}_{t,n}$
to denote an individual variable in $\boldsymbol{v}_{t}$. Let $\mathcal{H}=\left\{ n|\forall\boldsymbol{v}_{t,n}\in\boldsymbol{v}_{t}\right\} $.
Moreover, we use $q\left(\boldsymbol{v}_{t}\right)$ as a simplified
notation for $q\left(\boldsymbol{v}_{t}|\boldsymbol{y}_{1:t};\boldsymbol{\varphi}_{t}\right)$
when there is no ambiguity. The approximate conditional marginal posterior
$q\left(\boldsymbol{v}_{t}\right)$ could be calculated by minimizing
the Kullback-Leibler divergence (KLD) between $p\left(\boldsymbol{v}_{t}|\boldsymbol{y}_{1:t};\boldsymbol{\varphi}_{t}\right)$
and $q\left(\boldsymbol{v}_{t}\right)$ subject to a factorized form
constraint on $q\left(\boldsymbol{v}_{t}\right)$ as
\begin{align}
\mathcal{\mathscr{A}}_{\mathrm{VBI}}:\thinspace\thinspace\thinspace & q^{*}\left(\boldsymbol{v}_{t}\right)=\arg\min_{q\left(\boldsymbol{v}_{t}\right)}\int q\left(\boldsymbol{v}_{t}\right)\ln\frac{q\left(\boldsymbol{v}_{t}\right)}{p\left(\boldsymbol{v}_{t}|\boldsymbol{y}_{1:t};\boldsymbol{\varphi}_{t}\right)}d\boldsymbol{v}_{t}\label{eq:KLDmin}\\
 & \mathrm{s.t.}\thinspace\thinspace\thinspace\thinspace q\left(\boldsymbol{v}_{t}\right)=\prod_{n\in\mathcal{H}}q\left(\boldsymbol{v}_{t,n}\right),\label{eq:factorconstrain}\\
 & \thinspace\thinspace\thinspace\thinspace\thinspace\thinspace\thinspace\int q\left(\boldsymbol{v}_{t,n}\right)d\boldsymbol{v}_{t,n}=1,\forall n\in\mathcal{H}.
\end{align}
Problem $\mathcal{\mathscr{A}}_{\mathrm{VBI}}$ is non-convex, we
aim at finding a stationary solution (denoted by $q^{*}\left(\boldsymbol{v}_{t}\right)$)
of $\mathcal{\mathscr{A}}_{\mathrm{VBI}}$, as defined below.}
\begin{defn}
\textcolor{black}{[Stationary Solution]\label{lem:optimality-conditon-1}$q^{*}\left(\boldsymbol{v}_{t}\right)=\prod_{n\in\mathcal{H}}q^{*}\left(\boldsymbol{v}_{t,n}\right)$
is called a stationary solution of Problem $\mathcal{\mathscr{A}}_{\mathrm{VBI}}$
if it satisfies all the constraints in $\mathcal{\mathscr{A}}_{\mathrm{VBI}}$
and $\forall n\in\mathcal{H}$,
\begin{multline*}
q^{*}\left(\boldsymbol{v}_{t,n}\right)=\\
\arg\min_{q\left(\boldsymbol{v}_{t,n}\right)}\int\prod_{l\neq n}q^{*}\left(\boldsymbol{v}_{t,l}\right)q\left(\boldsymbol{v}_{t,n}\right)\ln\frac{\prod_{l\neq n}q^{*}\left(\boldsymbol{v}_{t,l}\right)q\left(\boldsymbol{v}_{t,n}\right)}{p\left(\boldsymbol{v}_{t}|\boldsymbol{y}_{1:t};\boldsymbol{\varphi}_{t}\right)}d\boldsymbol{v}_{t}.
\end{multline*}
}

\end{defn}
\textcolor{black}{By finding a stationary solution $q^{*}\left(\boldsymbol{v}_{t}\right)$
of $\mathcal{\mathscr{A}}_{\mathrm{VBI}}$, we could obtain the approximate
posterior $q^{*}\left(\boldsymbol{v}_{t,n}\right)\thickapprox p\left(\boldsymbol{v}_{t,n}|\boldsymbol{y}_{1:t};\boldsymbol{\varphi}_{t}\right),\forall n\in\mathcal{H}$. }

\textcolor{black}{A stationary solution of $\mathcal{\mathscr{A}}_{\mathrm{VBI}}$
can be obtained via alternately optimizing each individual density
$q\left(\boldsymbol{v}_{t,n}\right),n\in\mathcal{H}$. For given $q\left(\boldsymbol{v}_{t,l}\right),\forall l\neq n$,
the optimal $q\left(\boldsymbol{v}_{t,n}\right)$ that minimizes the
KLD in $\mathcal{\mathscr{A}}_{\mathrm{VBI}}$ is given by
\begin{equation}
q\left(\boldsymbol{v}_{t,n}\right)\propto\exp\left(\left\langle \ln p(\boldsymbol{y}_{1:t},\boldsymbol{v}_{t};\boldsymbol{\varphi}_{t})\right\rangle _{\prod_{l\neq n}q\left(\boldsymbol{v}_{t,l}\right)}\right),\label{eq:optimal_q}
\end{equation}
where $\left\langle f\left(x\right)\right\rangle _{q(x)}=\int f\left(x\right)q(x)dx$.
However, $p(\boldsymbol{y}_{1:t},\boldsymbol{v}_{t};\boldsymbol{\varphi}_{t})$
is intractable. Since $p(\boldsymbol{y}_{1:t},\boldsymbol{v}_{t};\boldsymbol{\varphi}_{t})\approx\hat{p}(\boldsymbol{y}_{1:t},\boldsymbol{v}_{t};\boldsymbol{\varphi}_{t})$
in (\ref{eq:pyvfappro}), \eqref{eq:optimal_q} can be approximated
as
\begin{equation}
q\left(\boldsymbol{v}_{t,n}\right)\propto\exp\left(\left\langle \ln\hat{p}(\boldsymbol{y}_{1:t},\boldsymbol{v}_{t};\boldsymbol{\varphi}_{t})\right\rangle _{\prod_{l\neq n}q\left(\boldsymbol{v}_{t,l}\right)}\right).\label{eq:optimal_q-1}
\end{equation}
Based on \eqref{eq:optimal_q-1}, the update equations of all variables
are given in the subsequent subsections. The detailed derivation can
be found in Appendix \ref{subsec:Derivation-1}. Note that the operator
$\left\langle \cdot\right\rangle _{\boldsymbol{v}_{t,l}}$ is equivalent
to $\left\langle \cdot\right\rangle _{q\left(\boldsymbol{v}_{t,l}\right)}$
and the expectation $\left\langle f\left(\boldsymbol{v}_{t,l}\right)\right\rangle _{q\left(\boldsymbol{v}_{t,l}\right)}$
w.r.t. its own approximate posterior is simplified as $\left\langle f\left(\boldsymbol{v}_{t,l}\right)\right\rangle $. }

\subsubsection{\textcolor{black}{Initialization of Sparse VBI\label{subsec:Initialization-of-Sparse}}}

\textcolor{black}{In order to trigger the alternating optimization
(AO) algorithm, we use the following initializations for the distribution
functions $q\left(\boldsymbol{x}_{t}\right),q(\boldsymbol{\gamma}_{t})$
in the first iteration of every frame and $q\left(\boldsymbol{s}_{1}\right)$
in the first iteration of first frame. In the rest iterations, we
initialize $q\left(\boldsymbol{x}_{t}\right)$,$q\left(\boldsymbol{s}_{t}\right)$,$q(\boldsymbol{\gamma}_{t})$
to the (approximate) posterior calculated in the previous frame.}
\begin{itemize}
\item \textcolor{black}{Initialize $q\left(\boldsymbol{s}_{1}\right)=\hat{p}(\boldsymbol{s}_{1}|\boldsymbol{y}_{1};\boldsymbol{\varphi}_{t})=\prod_{n=1}^{\tilde{N}}q\left(s_{1,n}\right)$
with $q\left(s_{1,n}\right)=\left(\tilde{\pi}_{1,n}\right)^{s_{1.n}}\left(1-\tilde{\pi}_{1,n}\right)^{1-s_{1,n}}$. }
\item \textcolor{black}{For given $\hat{p}(\boldsymbol{s}_{t}|\boldsymbol{y}_{1:t};\boldsymbol{\varphi}_{t})=\prod_{n=1}^{\tilde{N}}\left(\tilde{\pi}_{t,n}\right)^{s_{t.n}}\left(1-\tilde{\pi}_{t,n}\right)^{1-s_{t,n}}$,
initialize a gamma distribution for $\mathbf{\boldsymbol{\gamma}}_{t}$:
$q\left(\mathbf{\boldsymbol{\gamma}}_{t}\right)=\prod_{n=1}^{\tilde{N}}\Gamma\left(\gamma_{t,n};\tilde{a}_{\gamma,t,n},\tilde{b}_{\gamma,t,n}\right)$,
where $\tilde{a}_{\gamma,t,n}=\tilde{\pi}_{t,n}a_{t}+\left(1-\tilde{\pi}_{t,n}\right)\overline{a}_{t}$,
$\tilde{b}_{\gamma,t,n}=\tilde{\pi}_{t,n}b_{t}+\left(1-\tilde{\pi}_{t,n}\right)\overline{b}_{t}$.}
\item \textcolor{black}{Initialize a Gaussian distribution for $\boldsymbol{x}_{t}$:
$q\left(\boldsymbol{x}_{t}\right)=\mathcal{CN}(\boldsymbol{x}_{t};\boldsymbol{\mu}_{t},\boldsymbol{\Sigma}_{t})$
, where $\boldsymbol{\Sigma}_{t}=\left(\mathrm{diag}\left(\left\langle \mathbf{\boldsymbol{\gamma}}_{t}\right\rangle \right)+\left(\boldsymbol{F}_{t}\right)^{H}\boldsymbol{F}_{t}\right)^{-1},$
$\boldsymbol{\mu}_{t}=\boldsymbol{\Sigma}_{t}\left(\boldsymbol{F}_{t}\right)^{H}\boldsymbol{y}_{t}.$}
\end{itemize}

\subsubsection{\textcolor{black}{Update for $q\left(\kappa_{t}\right)$}}

\textcolor{black}{From (\ref{eq:optimal_q-1}), $q\left(\kappa_{t}\right)$
can be derived as
\begin{equation}
q\left(\kappa_{t}\right)=\Gamma(\kappa_{t};\tilde{a}_{\kappa,t},\tilde{b}_{\kappa,t}).\label{eq:poster_kappa}
\end{equation}
where $\tilde{a}_{\kappa,t}=a_{\kappa}+NN_{p}$, $\tilde{b}_{\kappa,t}=b_{\kappa}+\left\langle \left\Vert \boldsymbol{y}_{t}-\boldsymbol{F}_{t}\boldsymbol{x}_{t}\right\Vert ^{2}\right\rangle _{\boldsymbol{x}_{t}}=b_{\kappa}+\left\Vert \boldsymbol{y}_{t}-\boldsymbol{F}_{t}\boldsymbol{\mu}_{t}\right\Vert ^{2}+\mathrm{tr}\left(\boldsymbol{F}_{t}\boldsymbol{\Sigma}_{t}\left(\boldsymbol{F}_{t}\right)^{H}\right)$. }

\subsubsection{\textcolor{black}{Update for $q(\boldsymbol{x}_{t})$}}

\textcolor{black}{$q\left(\boldsymbol{x}_{t}\right)$ can be derived
as
\begin{equation}
q\left(\boldsymbol{x}_{t}\right)=\mathcal{CN}\left(\boldsymbol{x}_{t};\boldsymbol{\mu}_{t},\boldsymbol{\Sigma}_{t}\right).\label{eq:poster_x}
\end{equation}
$\boldsymbol{\mu}_{t}$ and $\boldsymbol{\Sigma}_{t}$ can be calculated
through
\begin{alignat}{1}
\boldsymbol{\Sigma}_{t} & =\left(\mathrm{diag}\left(\left\langle \mathbf{\boldsymbol{\gamma}}_{t}\right\rangle \right)+\left\langle \kappa_{t}\right\rangle \left(\boldsymbol{F}_{t}\right)^{H}\boldsymbol{F}_{t}\right)^{-1},\nonumber \\
{\color{black}{\color{black}}} & {\color{black}{\color{black}{\color{black}{\color{blue}=\mathbf{R}-\upsilon\mathbf{R}\left(\boldsymbol{F}_{t}\right)^{H}\left(\mathbf{I}+\upsilon\boldsymbol{F}_{t}\mathbf{R}\left(\boldsymbol{F}_{t}\right)^{H}\right)^{-1}\boldsymbol{F}_{t}\mathbf{R}.}}}}\label{eq:Sigma_x}
\end{alignat}
}

\textcolor{black}{
\begin{equation}
\boldsymbol{\mu}_{t}=\left\langle \kappa_{t}\right\rangle \boldsymbol{\Sigma}_{t}\left(\boldsymbol{F}_{t}\right)^{H}\boldsymbol{y}_{t}.\label{eq:Mu_x}
\end{equation}
}

\noindent \textcolor{black}{where $\left\langle \mathbf{\boldsymbol{\gamma}}_{t}\right\rangle =\frac{\tilde{a}_{\gamma,t,n}}{\tilde{b}_{\gamma,t,n}}$,
$\left\langle \kappa_{t}\right\rangle =\frac{\tilde{a}_{\kappa,t}}{\tilde{b}_{\kappa,t}}$,
$\mathbf{R}=\mathrm{diag}\left(\left[\frac{\tilde{b}_{\gamma,t,1}}{\tilde{a}_{\gamma,t,1}},\cdots,\frac{\tilde{b}_{\gamma,t,\tilde{N}}}{\tilde{a}_{\gamma,t,\tilde{N}}}\right]\right)$,
$\upsilon=\left\langle \kappa_{t}\right\rangle =\frac{\tilde{a}_{\kappa,t}}{\tilde{b}_{\kappa,t}}$.}

\subsubsection{\textcolor{black}{Update for $q\left(\mathbf{\boldsymbol{\gamma}}_{t}\right)$}}

\textcolor{black}{$q\left(\mathbf{\boldsymbol{\gamma}}_{t}\right)$
can be derived as}

\textcolor{black}{
\begin{equation}
q\left(\mathbf{\boldsymbol{\gamma}}_{t}\right)=\prod_{n=1}^{\tilde{N}}\Gamma\left(\gamma_{t,n};\tilde{a}_{\gamma,t,n},\tilde{b}_{\gamma,t,n}\right),\label{eq:poster_rho}
\end{equation}
where $\tilde{a}_{\gamma,t,n},\tilde{b}_{\gamma,t,n}$ are given by:
\begin{align}
\tilde{a}_{\gamma,t,n}= & \left\langle s_{t,n}\right\rangle a_{t}+\left\langle 1-s_{t,n}\right\rangle \overline{a}_{t}+1,\label{eq:a_tilde}\\
\tilde{b}_{\gamma,t,n}= & \left\langle s_{t,n}\right\rangle b_{t}+\left\langle 1-s_{t,n}\right\rangle \overline{b}_{t}+\left\langle \left|x_{t,n}\right|^{2}\right\rangle .\label{eq:b_tilde}
\end{align}
}

\noindent \textcolor{black}{where $\left\langle s_{t,n}\right\rangle =\tilde{\pi}_{t,n},\left\langle 1-s_{t,n}\right\rangle =1-\tilde{\pi}_{t,q},$
$\left\langle \left|x_{t,n}\right|^{2}\right\rangle =\left|\mu_{t,n}\right|^{2}+\Sigma_{t,n}$,
$\mu_{t,n}$ is the $n$-th element of $\boldsymbol{\mu}_{t}$, $\Sigma_{t,n}$
is the $n$-th diagonal element of $\boldsymbol{\Sigma}_{t}$.}

\subsubsection{\textcolor{black}{Update for $q\left(\boldsymbol{s}_{t}\right)$}}

\textcolor{black}{$q\left(\boldsymbol{s}_{t}\right)$ can be derived
as }

\textcolor{black}{
\begin{equation}
q\left(\boldsymbol{s}_{t}\right)=\prod_{n=1}^{\tilde{N}}\left(\pi_{t,n}\right)^{s_{t.n}}\left(1-\pi_{t,n}\right)^{1-s_{t,n}},\label{eq:poster_q}
\end{equation}
where $\pi_{t,n}$ is given by }

\textcolor{black}{
\begin{align}
\pi_{t,n}=\frac{1}{C} & \frac{\tilde{\pi}_{t,n}b_{t}^{a_{t}}}{\Gamma(a_{t})}e^{\left(a_{t}-1\right)\left\langle \ln\mathbf{\boldsymbol{\gamma}}_{t,n}\right\rangle -b_{t}\left\langle \mathbf{\boldsymbol{\gamma}}_{t,n}\right\rangle },\label{eq:pi_tilde}
\end{align}
and $C$ is the normalization constant, given by $C=\frac{\tilde{\pi}_{t,n}b_{t}^{a_{t}}}{\Gamma(a_{t})}e^{\left(a_{t}-1\right)\left\langle \ln\mathbf{\boldsymbol{\gamma}}_{t,n}\right\rangle -b_{t}\left\langle \mathbf{\boldsymbol{\gamma}}_{t,n}\right\rangle }+\frac{(1-\tilde{\pi}_{t,n})\overline{b}_{t}^{\overline{a}_{t}}}{\Gamma(\overline{a}_{t})}e^{\left(\overline{a}_{t}-1\right)\left\langle \ln\mathbf{\boldsymbol{\gamma}}_{t,n}\right\rangle -\overline{b}_{t}\left\langle \mathbf{\boldsymbol{\gamma}}_{t,n}\right\rangle }$,
$\left\langle \ln\gamma_{t,n}\right\rangle =\psi\left(\tilde{a}_{\gamma,t,n}\right)-\ln\left(\tilde{b}_{\gamma,t,n}\right)$,
$\psi\left(x\right)=\frac{d}{dx}\ln\left(\Gamma\left(x\right)\right)$
is the digamma function, defined as the logarithmic derivative of
the gamma function.}

\textcolor{black}{}
\begin{algorithm}
\textcolor{black}{\caption{\label{alg1}Doppler-Aware-Dynamic-VBI algorithm}
}

\textcolor{black}{1: }\textbf{\textcolor{black}{for}}\textcolor{black}{{}
$t=1,2,...$ }\textbf{\textcolor{black}{do}}

\textcolor{black}{2: ~~Initialize the distribution functions according
to Section }

\textcolor{black}{~~~~~~\ref{subsec:Initialization-of-Sparse}.}

\textcolor{black}{3: ~~}\textbf{\textcolor{black}{while}}\textcolor{black}{{}
not converge }\textbf{\textcolor{black}{do}}

\textcolor{black}{4: ~~}\textbf{\textcolor{black}{while}}\textcolor{black}{{}
not converge }\textbf{\textcolor{black}{do}}

\textcolor{black}{5: ~~~~}\textbf{\textcolor{black}{{} \%DD-VBI-E
Step: }}

\textcolor{black}{6: ~~~~ Update $q\left(\kappa_{t}\right)$ using
(\ref{eq:poster_kappa}).}

\textcolor{black}{7: ~~~~ Update $q(\boldsymbol{x}_{t})$ using
(\ref{eq:poster_x}).}

\textcolor{black}{8: ~~~~ Update $q\left(\mathbf{\boldsymbol{\gamma}}_{t}\right)$
using (\ref{eq:poster_rho}).}

\textcolor{black}{9: ~~~~ Update $q\left(\boldsymbol{s}_{t}\right)$
using (\ref{eq:poster_q}) .}

\textcolor{black}{10: ~~}\textbf{\textcolor{black}{end while}}

\textcolor{black}{11: ~~~~}\textbf{\textcolor{black}{{} \%DD-VBI-M
Step: }}

\textcolor{black}{12: ~~~~ Construct the surrogate function $\hat{u}$
in (\ref{eq:surr_1}) using the }

\textcolor{black}{~~~~~~~~~ output of DD-VBI-E Step $q(\boldsymbol{v}_{t}|\boldsymbol{y}_{1:t};\boldsymbol{\varphi}_{t}).$}

\textcolor{black}{13: ~~~~ Update $\boldsymbol{\varphi}_{t}$
using (\ref{eq:max_surr}).}

\textcolor{black}{14: ~~}\textbf{\textcolor{black}{end while}}

\textcolor{black}{15: ~~Let $\hat{f}_{d,t}$ denote the converged
maximum DFO for frame}

\textcolor{black}{~~~~~~~$t$. Calculate $\hat{p}(\boldsymbol{s}_{t+1}|\boldsymbol{y}_{1:t};\boldsymbol{\hat{\varphi}}_{1:t})$
using (\ref{eq:Pheads}).}

\textcolor{black}{~~~~~~~Pass messages $\hat{p}(\boldsymbol{s}_{t+1}|\boldsymbol{y}_{1:t};\boldsymbol{\hat{\varphi}}_{1:t})$
to frame $t+1$.}

\textcolor{black}{16: ~~Estimate ${x_{t,n}}$ using (\ref{eq:Mu_x}).}

\textcolor{black}{17: ~~}\textbf{\textcolor{black}{end for}}
\end{algorithm}

\subsection{\textcolor{black}{Messages $\hat{p}(\boldsymbol{s}_{t+1}|\boldsymbol{y}_{1:t};\boldsymbol{\hat{\varphi}}_{1:t})$
Passed to the Next Frame \label{subsec:Cauculating--passed}}}

\textcolor{black}{After the convergence of the DD-VBI iterations in
frame $t$, let $\hat{f}_{d,t}$ and $q(\boldsymbol{s}_{t}|\boldsymbol{y}_{1:t},;\boldsymbol{\hat{\varphi}}_{1:t})=\prod_{n=1}^{\tilde{N}}\left(\pi_{t,n}\right)^{s_{t.n}}\left(1-\pi_{t,n}\right)^{1-s_{t,n}}$
denote the converged Doppler parameter and the associated approximate
posterior for $\boldsymbol{s}_{t}$, respectively. Then, the messages
$\hat{p}(\boldsymbol{s}_{t+1}|\boldsymbol{y}_{1:t};;\boldsymbol{\hat{\varphi}}_{1:t})$
is calculated as}

\textcolor{black}{
\begin{align}
 & \hat{p}(\boldsymbol{s}_{t+1}|\boldsymbol{y}_{1:t};\boldsymbol{\hat{\varphi}}_{1:t})\nonumber \\
= & \prod_{n=1}^{\tilde{N}}\sum_{\mathbf{s}_{t,n}}q\left(s_{t,n}|\boldsymbol{y}_{1:t},\boldsymbol{\hat{\varphi}}_{1:t}\right)p(s_{t+1,n}|s_{t,n})\\
= & \prod_{n=1}^{\tilde{N}}\left(\tilde{\pi}_{t+1,n}\right)^{s_{t+1,n}}\left(1-\tilde{\pi}_{t+1,n}\right)^{1-s_{t+1,n}}.\label{eq:Pheads}
\end{align}
where}

\textcolor{black}{
\[
\tilde{\pi}_{t+1,n}=\left(1-\pi_{t,n}\right)\rho_{0,1}+\pi_{t,n}(1-\rho_{1,0}).
\]
Finally, the messages $\hat{p}(\boldsymbol{s}_{t+1}|\boldsymbol{y}_{1:t};\boldsymbol{\hat{\varphi}}_{1:t})$,
as the prior to the channel support, is passed to the next frame.}

\textcolor{black}{The overall DD-VBI algorithm is summarized in Algorithm
1. Note that in the $t$-th frame of DD-VBI, the contribution of the
previous observations $\boldsymbol{y}_{1:t-1}$ on the estimation
of $\boldsymbol{v}_{t}$ and $\boldsymbol{\varphi}_{t}$ is summarized
in the messages $\hat{p}(\boldsymbol{s}_{t}|\boldsymbol{y}_{1:t-1};\boldsymbol{\hat{\varphi}}_{1:t-1})$
passed from frame $t-1$. Therefore, in the $t$-th frame, there is
no need to store all the observations $\boldsymbol{y}_{1:t-1}$ up
to frame $t-1$.}
\begin{rem}
\textcolor{black}{In practical mmWave massive MIMO systems, the number
of RF chains can be less than the number of antennas at both the BS
and user sides to reduce the hardware cost and power consumption.
The proposed scheme can be easily extended to the case with limited
RF chains. For example, suppose there are only $M_{b}<M$ RF chains
at the BS and $N_{b}<N$ RF chains at the user. In this case, when
the BS transmits the training vector $\mathbf{v}=\mathbf{g}\mathbf{F}\in\mathbb{C}^{M}$
for downlink channel tracking in the $i$-th symbol duration, the
user employs $\mathbf{U}_{i}=\mathbf{W}_{i}\mathbf{G}_{i}\in\mathbb{C}^{N\times N_{b}}$
as a combining matrix to combine the received signal into $N_{b}$
baseband channel measurements, where $\mathbf{F}\in\mathbb{C}^{M\times M_{b}}$
and $\mathbf{g}\in\mathbb{C}^{M_{b}}$ are the RF training matrix
and baseband training vector at the BS, respectively, and $\mathbf{W}_{i}\in\mathbb{C}^{N\times N_{b}}$
and $\mathbf{G}_{i}\in\mathbb{C}^{N_{b}\times N_{b}}$ are the RF
and baseband combining matrix at the mobile user in the $i$-th symbol
duration, respectively. We can still write the received pilot signal
as a CS model as in (\ref{eq:compress formulation}), but with $\boldsymbol{F}_{t}=[\boldsymbol{F}_{t,1};...;\boldsymbol{F}_{t,N_{p}}]\text{\ensuremath{\in}}C^{N_{b}N_{p}\text{\texttimes}\tilde{N}}$,
$\boldsymbol{F}_{t,i}=\boldsymbol{U}_{t,i}^{H}\boldsymbol{A}_{R,i}(\boldsymbol{\varphi}_{t})$,
$\boldsymbol{n}_{t}=[\boldsymbol{U}_{t,i}^{H}\boldsymbol{n}_{t,i}]_{i\in\mathcal{N}_{p}}$.}
\end{rem}

\subsection{\textcolor{black}{Complexity and Signaling Overhead Comparison\label{subsec:Complexity-issue-and} }}

\textcolor{black}{The computational complexity of the proposed algorithm
is dominated by the update of $q(\boldsymbol{x}_{t})$. Assuming the
arithmetic with individual elements has complexity $\mathcal{O}(1)$,
the computational complexity of matrix inversion in (\ref{eq:Sigma_x})
is $\mathcal{O}(N_{b}^{3}N_{p}^{3})$ and the total number of multiplications
to update $q(\boldsymbol{x}_{t})$ is $3\tilde{N}+2N_{b}N_{p}\tilde{N}^{2}+2(N_{b}N_{p})^{2}\tilde{N}+(N_{b}N_{p})^{2}$.
Supposing the algorithm executes $I$ iterations, the total complexity
order of the proposed method is $\mathcal{O}\left(I(N_{b}N_{p})^{2}\tilde{N}\right)$,
considering that $\tilde{N}$ is usually larger than $N_{b}N_{p}$.
In Table \ref{tab:Asymptotic-complexity-1} and \ref{tab:Asymptotic-complexity-1-2-1-2},
we assume the mmWave channel has $L_{D}$ dominant paths and compare
the complexity and signaling overhead of the proposed algorithm with
the following baseline algorithms:}
\begin{itemize}
\item \textbf{\textcolor{black}{Baseline 1}}\textcolor{black}{{} (ML) \cite{8558718}:
This is the ML based joint DFO and channel estimation algorithm proposed
in \cite{8558718}. }
\item \textbf{\textcolor{black}{Baseline 2}}\textcolor{black}{{} (ES) \cite{Wang2009Beam}:
This is the codebook based exhaustive beam search algorithm in \cite{Wang2009Beam}.
$K_{E}$ denotes the number of beamforming vectors of the codebook
in ES.}
\item \textbf{\textcolor{black}{Baseline 3}}\textcolor{black}{{} (HS) \cite{Alkhateeb2017Channel}:
This is the codebook based hierarchical beam search algorithm in \cite{Alkhateeb2017Channel}.
$S$ denotes the total level of hierarchical codebook. $K_{H}$ denotes
the number of beamforming vectors of each codebook level in HS.}
\end{itemize}
\textcolor{black}{As seen from Table \ref{tab:Asymptotic-complexity-1},
the complexity order of the proposed algorithm is similar to the baselines.
For example, $N_{b}N_{p}$ can range from $\mathcal{O}\left(L_{D}\right)$
to $\mathcal{O}\left(N\right)$ to achieve different tradeoff between
performance and complexity, and we usually have $\tilde{N}=\mathcal{O}\left(N\right)$,
$K_{E}=\mathcal{O}\left(N\right)$. For a resolution $\mathcal{O}(2\pi/N)$
of the HS scheme, we usually have $\mathcal{S=O}\left(\log_{K_{H}}(N/L_{D})\right)$
\cite{Alkhateeb2017Channel}. In this typical case, the complexity
of the proposed scheme and the baseline schemes are shown in the third
column of Table \ref{tab:Asymptotic-complexity-1}. On the other hand,
the proposed algorithm has the lowest signaling overhead for both
the general case and the typical case, as shown in Table \ref{tab:Asymptotic-complexity-1-2-1-2}.}

\textcolor{black}{}
\begin{table}
\begin{centering}
\textcolor{black}{}%
\begin{tabular}{|c|c|c|}
\hline 
\textcolor{black}{Algorithms} & \textcolor{black}{Complexity order} & \textcolor{black}{Typical complexity order}\tabularnewline
\hline 
\hline 
\textcolor{black}{proposed} & \textcolor{black}{$\mathcal{O}\left(I(N_{b}N_{p})\tilde{N}^{2}\right)$} & \textcolor{black}{$\mathcal{O}\left(IL_{D}^{2}N\right)-\mathcal{O}\left(IN^{3}\right)$}\tabularnewline
\hline 
\textcolor{black}{ML} & \textcolor{black}{$\mathcal{O}\left(\tilde{N}^{3}\right)$} & \textcolor{black}{$\mathcal{O}\left(N^{3}\right)$}\tabularnewline
\hline 
\textcolor{black}{ES} & \textcolor{black}{$\mathcal{O}\left(L_{D}^{2}K_{E}^{4}\right)$} & \textcolor{black}{$\mathcal{O}\left(L_{D}^{2}N^{4}\right)$}\tabularnewline
\hline 
\textcolor{black}{HS} & \textcolor{black}{$\mathcal{O}\left(SL_{D}^{4}K_{H}^{2}\right)$} & \textcolor{black}{$\mathcal{O}\left(\log_{K_{H}}(N/L_{D})L_{D}^{4}K_{H}^{2}\right)$}\tabularnewline
\hline 
\end{tabular}
\par\end{centering}
\textcolor{black}{\caption{\textcolor{blue}{\label{tab:Asymptotic-complexity-1}}\textcolor{black}{Complexity
orders for different schemes.}}
}
\end{table}

\textcolor{black}{}
\begin{table}
\begin{centering}
\textcolor{black}{}%
\begin{tabular}{|c|c|c|}
\hline 
\textcolor{black}{Algorithms} & \textcolor{black}{Signaling overhead} & \textcolor{black}{Typical signaling overhead}\tabularnewline
\hline 
\hline 
\textcolor{black}{proposed} & \textcolor{black}{$\mathcal{O}(L_{D}/N_{b})+L_{D}$} & \textcolor{black}{$4+L_{D}$}\tabularnewline
\hline 
\textcolor{black}{ML} & \textcolor{black}{$N$} & \textcolor{black}{$N$}\tabularnewline
\hline 
\textcolor{black}{HS} & \textcolor{black}{$K_{H}^{2}\left(L_{D}\right)^{3}S$} & \textcolor{black}{$K_{H}^{2}\left(L_{D}\right)^{3}\log_{K_{H}}(N/L_{D})$}\tabularnewline
\hline 
\textcolor{black}{ES} & \textcolor{black}{$K_{E}^{2}$} & \textcolor{black}{$N^{2}$}\tabularnewline
\hline 
\end{tabular}
\par\end{centering}
\textcolor{black}{\caption{\textcolor{blue}{\label{tab:Asymptotic-complexity-1-2-1-2}}\textcolor{black}{Signaling
overhead for different schemes.}}
}
\end{table}

\section{\textcolor{black}{Simulation Results\label{sec:Simulation-Results}}}

\textcolor{black}{In this section, we compare the performance of the
proposed algorithm with the baseline algorithms described in Section
\ref{subsec:Complexity-issue-and}. For the ML baseline, we also consider
the case when only partial channel parameters are estimated as in
the proposed scheme (ML-Partial). For the proposed scheme, we also
consider the case when the training vector is generated randomly (DD-VBI-Random).
The channel parameters are based on the millimeter-wave statistical
spatial channel model (mm-SSCM) as specified in \cite{1Samimi120163},
which was developed according to the 28- and 73-GHz ultrawideband
propagation measurements in New York City. The signal bandwidth is
$50MHz$ and the frame duration is set as $T_{b}=0.5ms$. The carrier
frequency is 28GHz . The mobile user employs a ULA of $M=128$ antennas
and the inter antenna spacing is $\lambda/2$, and the BS also employs
a ULA. The user velocity is assumed to be 380km/h, which translates
to $f_{d,t}\approx10$KHz. }

\textcolor{black}{In the simulation, we will consider both cases when
the user is equipped with a full set of RF chains and limited RF chains.
For the case with limited RF chains, the number of RF chains is set
to be $N_{b}=16$. The MSE for DFO estimation and the uplink achievable
data rate are adopted as the performance metrics. The frequency MSE
and the channel MSE is defined as $\frac{\left\Vert \hat{f}_{d,t}-f_{d,t}\right\Vert ^{2}}{\left\Vert f_{d,t}\right\Vert ^{2}}$
and $\frac{\left\Vert \hat{\boldsymbol{x}}_{t}-\boldsymbol{x}_{t}\right\Vert ^{2}}{\left\Vert \boldsymbol{x}_{t}\right\Vert ^{2}}$,
respectively. The parameter $N_{d}$ is chosen to be equal to the
number of dominant AoA directions at the user side (an AoA direction
is called a dominant AoA direction if its energy is no less than 10\%
of the most significant AoA direction) and $N_{d}$ data streams are
transmitted over the $N_{d}$ dominant AoA directions in the uplink
with equal power allocation. Since the uplink transmission is only
designed based on the dominant AoA directions which are known at the
user, there is no need for the BS to feed back the slow time-varying
effective channel.}

\subsection{\textcolor{black}{Doppler Frequency and Channel MSE Performance}}

\begin{figure}[!tbph]
\begin{centering}
\textsf{\includegraphics[scale=0.5]{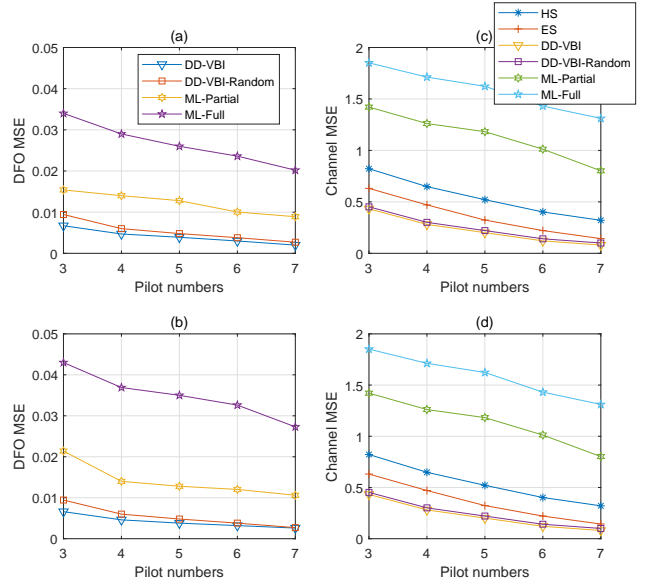}}
\par\end{centering}
\caption{\textcolor{blue}{\label{fig:Frequency-MSE-and}}Frequency MSE and
channel MSE performance versus the pilot number. Set SNR= 0 dB. (a)
full RF chains with $N=128$. (b) limited RF chains with $N=128$,
$N_{b}=16$\textcolor{black}{.}}
\end{figure}

\begin{figure}[!tbph]
\begin{centering}
\textsf{\textcolor{black}{\includegraphics[scale=0.5]{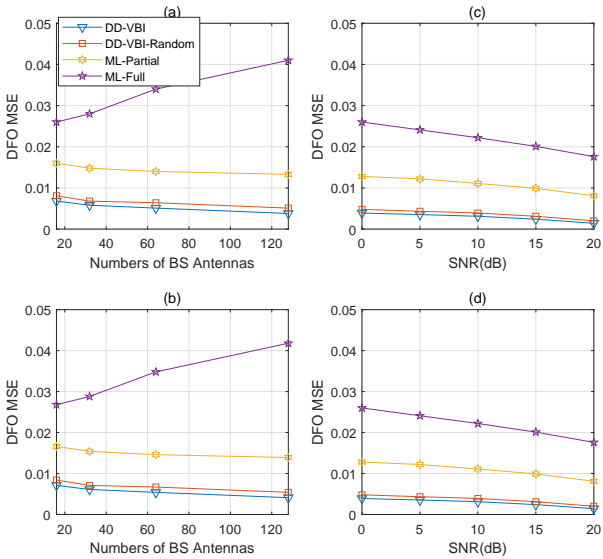}}}
\par\end{centering}
\textcolor{black}{\caption{\textcolor{black}{\label{fig:Frequency-MSE-performance}}Frequency
MSE performance versus the number of BS antennas and SNR. (a) Set
SNR= 0 dB. Full RF chains with $N=128$. (b) Set SNR= 0 dB. Limited
RF chains with $N=128$, $N_{b}=16$. (c) The number of pilots is
fixed as 5. Full RF chains with $N=128$. (d) The number of pilots
is fixed as 5. Limited RF chains with $N=128$, $N_{b}=16$.}
}
\end{figure}

\textcolor{black}{The Doppler frequency and channel MSE performance
of different algorithms versus the pilot number, SNR and number of
BS antennas are shown in Fig. \ref{fig:Frequency-MSE-and} and Fig.
\ref{fig:Frequency-MSE-performance}. It can be seen that the proposed
DD-VBI algorithm achieves large performance gain over all the baseline
algorithms, under both full RF chains and limited RF chains. By using
the proposed training vector design to strike a balance between }\textit{\textcolor{black}{exploitation}}\textcolor{black}{{}
of known channel directions and }\textit{\textcolor{black}{exploration}}\textcolor{black}{{}
of unknown channel directions, the DD-VBI algorithm could further
improve the MSE performance compared to the case with a random training
vector. This demonstrates that the proposed algorithm can effectively
estimate the maximum DFO by selective channel tracking and efficient
training vector design. Note that as the number of BS antennas increases,
the parameters to be estimated increase significantly and the spatial
resolution and array gain will also increase. Since the ML-Full method
does not exploit the selective channel estimation method or channel
sparsity to reduce the number of free parameters, its performance
may degrade with the number of BS antennas. However, the performance
gain of the proposed algorithm improves because in this case, the
number of parameters to be estimated does not increase with the number
of BS antennas. This demonstrates that the proposed algorithm is a
powerful method for accurate DFO estimation, even when both BS and
mobile user are equipped with massive MIMO, and the number of RF chains
at the user side is limited. }

\textcolor{black}{}

\subsection{\textcolor{black}{Achievable Data Rate Performance}}

\textcolor{black}{}
\begin{figure}[tp]
\begin{centering}
\textsf{\textcolor{black}{\includegraphics[scale=0.5]{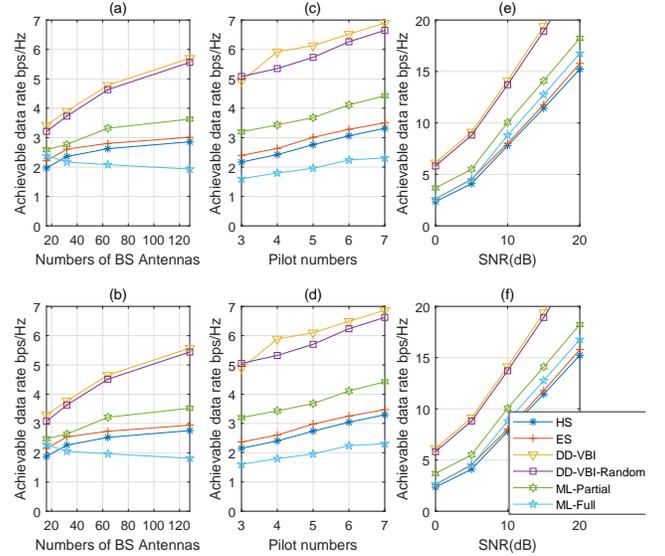}}}
\par\end{centering}
\textcolor{black}{\caption{\textcolor{blue}{\label{fig:Achievable-data-rate-1}}Achievable data
rate versus the number of BS antennas, pilot number and SNR. (a) and
(c) Set SNR= 0 dB. Full RF chains with $N=128$. (b) and (d) Set SNR=
0 dB. Limited RF chains with $N=128$, $N_{b}=16$. (e) The number
of pilots is fixed as 5. Full RF chains with $N=128$. (f) The number
of pilots is fixed as 5. Limited RF chains with $N=128$, $N_{b}=16$.}
}
\end{figure}

\textcolor{black}{The achievable data rates of different algorithms
versus the pilot number, SNR and number of BS antennas are shown in
Fig. \ref{fig:Achievable-data-rate-1}. It can be observed that the
performance of all algorithms increases with the number of pilots,
SNR and number of BS antennas. The proposed DD-VBI algorithm can achieve
large performance gain over various baselines under both full RF chains
and limited RF chains. Moreover, by using the proposed training vector
design, the proposed DD-VBI algorithm could further improve the achievable
data rate. This verifies that the proposed selective channel tracking
and Doppler compensation scheme can also enhance the achievable data
rate with low pilot overhead, under different SNRs and numbers of
BS antennas.}

\subsection{\textcolor{black}{Complexity versus Realized Gain}}

\begin{figure}[htbp]
\begin{centering}
\textsf{\textcolor{black}{\includegraphics[scale=0.5]{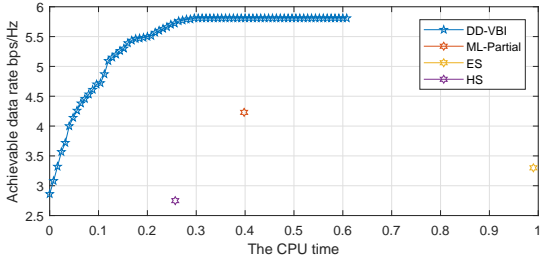}}}
\par\end{centering}
\textcolor{black}{\caption{\textcolor{blue}{\label{fig:The-tradeoff-between}}\textcolor{black}{The
tradeoff between the complexity and realized gain of the proposed
algorithm.}}
}
\end{figure}

\textcolor{black}{In practice, we can control the tradeoff between
the complexity and realized gain of the proposed algorithm by adjusting
the number of iterations. In Fig. \ref{fig:The-tradeoff-between},
we plot the achievable rate versus the CPU time. It can be seen that
the proposed algorithm can achieve a better performance than the baseline
algorithms for the same CPU time. Moreover, the performance gain increases
with the CPU time. Therefore, the proposed algorithm provides a better
and more flexible tradeoff between the performance and computational
power.}

\textcolor{black}{}

\section{\textcolor{black}{Conclusion\label{sec:Conclusion}}}

\textcolor{black}{We propose an angular-domain selective channel tracking
and Doppler compensation scheme for high-mobility massive MIMO systems.
Firstly, we propose a selective channel tracking scheme and the associated
Doppler-aware-dynamic-VBI algorithm to accurately estimate the DFO
and partial angular--domain channel parameters with reduced pilot
overhead. Then, we propose an angular-domain selective DFO compensation
scheme to convert the dominant paths of the fast time-varying channel
into a slow time-varying effective channel, based on which efficient
uplink and downlink transmissions can be achieved. Simulations verify
that the proposed scheme not only can mitigate the Doppler and channel
aging effect with much less pilots than existing schemes, but also
can achieve a good tradeoff between the CSI signaling overhead and
spatial multiplexing/array gain. }

\appendix

\subsection{\textcolor{black}{EM Estimation of the Parameters $\rho_{1,0}$ and
$\rho_{0,1}$ \label{subsec:Derivation-2}}}

\textcolor{black}{Denote the $\boldsymbol{s}_{t}$ as the channel
support vector, as defined in SubSec.\ref{subsec:Three-layer-Hierarchical-Markov}.
We apply the EM method in \cite{A1977Maximum} to obtain an estimate
of the parameter $\rho_{1,0}$ by perform the following iterations
until convergence:
\begin{equation}
\rho_{1,0}^{i+1}=\arg\max_{\rho_{1,0}}E_{\hat{p}^{i}(\boldsymbol{s}_{1:t})}\left\{ \ln p(\boldsymbol{s}_{1:t},\boldsymbol{y}_{1:t};\rho_{1,0})|\rho_{1,0}^{i}\right\} ,\label{eq:EM1}
\end{equation}
where $\rho_{1,0}^{i}$ stands for the $i$-th iteration, $E_{\hat{p}^{i}(\boldsymbol{s}_{1:t})}\left\{ \cdot\right\} $
denotes expectation over the posterior distribution $\hat{p}^{i}(\boldsymbol{s}_{1:t})=p(\boldsymbol{s}_{1:t}|\boldsymbol{y}_{1:t},\rho_{1,0}^{i})$
conditioned on $\boldsymbol{y}_{1:t}$ and $\rho_{1,0}^{i}$. Solving
the problem (\ref{eq:EM1}), we can get a closed-form expression for
updating the $\rho_{1,0}$ as
\begin{equation}
\rho_{1,0}^{i+1}=\frac{\sum_{\tilde{t}=2}^{t}\sum_{n=1}^{\tilde{N}}E_{\hat{p}^{i}(\boldsymbol{s}_{1:t})}\left[s_{\tilde{t},n}\right]-E_{\hat{p}^{i}(\boldsymbol{s}_{1:t})}\left[s_{\tilde{t}-1,n}s_{\tilde{t},n}\right]}{\sum_{\tilde{t}=2}^{t}\sum_{n=1}^{\tilde{N}}E_{\hat{p}^{i}(\boldsymbol{s}_{1:t})}\left[s_{\tilde{t}-1,n}\right]}\label{eq:rho}
\end{equation}
}

\noindent \textcolor{black}{Since the channel support vector $\boldsymbol{s}_{1:t}$
forms a Markov chain, the posterior distribution $\hat{p}^{i}(\boldsymbol{s}_{1:t})$
can be approximately calculated using the sum-product message passing
over the factor graph of the associated Markov chain with the priors
of $s_{\tilde{t},n}$'s given by the output of the DD-VBI algorithm
at each frame $\tilde{t}$ \cite{Ziniel2013Dynamic}. After the iterations
in (\ref{eq:rho}) convergence, we can obtain the EM estimation of
the parameter $\rho_{1,0}$ in the t-th frame. The EM estimation of
the parameter $\rho_{0,1}$ can be obtained in a similar way.}

\subsection{\textcolor{black}{Derivation of (\ref{eq:poster_kappa})-(\ref{eq:poster_q})
\label{subsec:Derivation-1}}}

\textcolor{black}{From (\ref{eq:optimal_q-1}), $q\left(\kappa_{t}\right)$
in (\ref{eq:poster_kappa}) can be obtained a
\begin{align*}
 & \ln q\left(\kappa_{t}\right)\wasypropto\left\langle \ln p(\boldsymbol{y}_{t}|\boldsymbol{x}_{t},\kappa_{t};\boldsymbol{\varphi}_{t})\right\rangle _{\boldsymbol{x}_{t}}+\ln p\left(\kappa_{t}\right)\\
\propto & -\kappa_{t}\left\langle \left\Vert \boldsymbol{y}_{t}-\boldsymbol{F}_{t}\boldsymbol{x}_{t}\right\Vert ^{2}\right\rangle _{\boldsymbol{x}_{t}}\\
 & +NN_{p}\ln\kappa_{t,i}+\left(a_{\kappa}-1\right)\ln\kappa_{t}-b_{\kappa}\kappa_{t}\\
\propto & \left(\tilde{a}_{\kappa}-1\right)\ln\kappa-\tilde{b}_{\kappa}\kappa.
\end{align*}
$q\left(\boldsymbol{x}_{t}\right)$ in (\ref{eq:poster_x}) can be
obtained as
\begin{align*}
 & \ln q\left(\boldsymbol{x}_{t}\right)\propto\left\langle \ln p\left(\boldsymbol{y}_{t}|\boldsymbol{x}_{t},\kappa_{t};\boldsymbol{\varphi}_{t}\right)\right\rangle _{\kappa_{t}}+\left\langle \ln p(\boldsymbol{x}_{t}|\mathbf{\boldsymbol{\gamma}}_{t})\right\rangle _{\mathbf{\boldsymbol{\gamma}}_{t}}\\
\propto & -\left\langle \kappa_{t}\right\rangle \left\Vert \boldsymbol{y}_{t}-\boldsymbol{F}_{t}\boldsymbol{x}_{t}\right\Vert ^{2}-\boldsymbol{x}_{t}^{H}\mathrm{diag}\left(\left\langle \mathbf{\boldsymbol{\gamma}}_{t}\right\rangle \right)\boldsymbol{x}_{t}\\
\propto & -\left(\boldsymbol{x}_{t}-\boldsymbol{\mu}_{t}\right)^{H}\left(\boldsymbol{\Sigma}_{t}\right)^{-1}\left(\boldsymbol{x}_{t}-\boldsymbol{\mu}_{t}\right).
\end{align*}
$q\left(\mathbf{\boldsymbol{\gamma}}_{t}\right)$ in (\ref{eq:poster_rho})
can be obtained as
\begin{align*}
 & \ln q\left(\mathbf{\boldsymbol{\gamma}}_{t}\right)\wasypropto\left\langle \ln p(\boldsymbol{x}_{t}|\mathbf{\boldsymbol{\gamma}}_{t})\right\rangle _{\boldsymbol{x}_{t}}+\left\langle \ln p\left(\mathbf{\boldsymbol{\gamma}}_{t}|\boldsymbol{s}_{t}\right)\right\rangle _{\boldsymbol{s}_{t}}\\
\wasypropto & \sum_{n}\left(\tilde{a}_{\gamma,t,n}-1\right)\ln\mathbf{\boldsymbol{\gamma}}_{t,n}-\tilde{b}_{\gamma,t,n}\mathbf{\boldsymbol{\gamma}}_{t,n}.
\end{align*}
$q\left(\boldsymbol{s}_{t}\right)$ in (\ref{eq:poster_q}) can be
obtained as
\begin{align*}
 & \wasypropto\ln q\left(\boldsymbol{s}_{t}\right)\wasypropto\left\langle \ln p\left(\mathbf{\boldsymbol{\gamma}}_{t}|\boldsymbol{s}_{t}\right)\right\rangle _{\boldsymbol{\gamma}_{t}}+\ln\hat{p}\left(\boldsymbol{s}_{t}\right)\\
\wasypropto & \sum_{n}s_{t}\left(\ln\frac{b_{t}^{a_{t}}}{\Gamma(a_{t})}+\left(a_{t}-1\right)\left\langle \ln\mathbf{\boldsymbol{\gamma}}_{t,n}\right\rangle \right.\\
 & \left.-b_{t}\left\langle \mathbf{\boldsymbol{\gamma}}_{t,n}\right\rangle +\ln\tilde{\pi}_{t,n}\right)+\left(1-s_{t}\right)\left(\ln\frac{\bar{b}_{t}^{\bar{a}_{t}}}{\Gamma(\bar{a}_{t})}+\left(\bar{a}_{t}-1\right)\right.\\
 & \left.\times\left\langle \ln\mathbf{\boldsymbol{\gamma}}_{t,n}\right\rangle -\bar{b}_{t}\left\langle \mathbf{\boldsymbol{\gamma}}_{t,n}\right\rangle +\ln\left(1-\tilde{\pi}_{t,n}\right)\right)\\
\wasypropto & \ln\prod_{n=1}^{\tilde{N}}\left(\pi_{t,n}\right)^{s_{t.n}}\left(1-\pi_{t,n}\right)^{1-s_{t,n}}.
\end{align*}
}



\begin{thebibliography}{10}
\providecommand{\url}[1]{#1}
\csname url@samestyle\endcsname
\providecommand{\newblock}{\relax}
\providecommand{\bibinfo}[2]{#2}
\providecommand{\BIBentrySTDinterwordspacing}{\spaceskip=0pt\relax}
\providecommand{\BIBentryALTinterwordstretchfactor}{4}
\providecommand{\BIBentryALTinterwordspacing}{\spaceskip=\fontdimen2\font plus
\BIBentryALTinterwordstretchfactor\fontdimen3\font minus
  \fontdimen4\font\relax}
\providecommand{\BIBforeignlanguage}[2]{{%
\expandafter\ifx\csname l@#1\endcsname\relax
\typeout{** WARNING: IEEEtran.bst: No hyphenation pattern has been}%
\typeout{** loaded for the language `#1'. Using the pattern for}%
\typeout{** the default language instead.}%
\else
\language=\csname l@#1\endcsname
\fi
#2}}
\providecommand{\BIBdecl}{\relax}
\BIBdecl

\bibitem{Souden2009Robust}
M.~Souden, S.~Affes, J.~Benesty, and R.~Bahroun, ``{Robust Doppler Spread
  Estimation in the Presence of a Residual Carrier Frequency Offset},''
  \emph{IEEE Transactions on Signal Processing}, vol.~57, no.~10, pp.
  4148--4153, 2009.

\bibitem{Berger2010Application}
C.~R. Berger, Z.~Wang, J.~Huang, and S.~Zhou, ``{Application of compressive
  sensing to sparse channel estimation},'' \emph{IEEE Communications Magazine},
  vol.~48, no.~11, pp. 164--174, 2010.

\bibitem{Wang2018Channel}
X.~Wang, G.~Wang, R.~Fan, and B.~Ai, ``{Channel Estimation with Expectation
  Maximization and Historical Information based Basis Expansion Model for
  Wireless Communication Systems on High Speed Railways},'' \emph{IEEE Access},
  vol.~6, no.~99, pp. 72--80, 2018.

\bibitem{Mostofi2005ICI}
Y.~Mostofi and D.~C. Cox, ``{ICI Mitigation for Pilot-aided OFDM Mobile
  Systems},'' \emph{IEEE Transactions on Wireless Communications}, vol.~4,
  no.~2, pp. 765--774, 2005.

\bibitem{Liu2015On}
T.~L. Liu, W.~H. Chung, S.~Y. Yuan, and S.~Y. Kuo, ``{On the Banded
  Approximation of the Channel Matrix for Mobile OFDM Systems},''
  \emph{Vehicular Technology IEEE Transactions on}, vol.~64, no.~8, pp.
  3526--3535, 2015.

\bibitem{Xin2014Study}
W.~Xin, X.~X. Yang, and Y.~Z. Wang, ``{Study on Channel Estimation in
  Time-Varing Amplify and Forward Relay Networks},'' \emph{Applied Mechanics
  Materials}, vol. 519-520, pp. 934--938, 2014.

\bibitem{Al2010A}
T.~Y. Al-Naffouri, K.~M.~Z. Islam, N.~Al-Dhahir, and S.~Lu, ``{A Model
  Reduction Approach for OFDM Channel Estimation Under High Mobility
  Conditions},'' \emph{IEEE Transactions on Signal Processing}, vol.~58, no.~4,
  pp. 2181--2193, 2010.

\bibitem{8558718}
Y.~Ge, W.~Zhang, F.~Gao, and H.~Minn, ``{Angle-Domain Approach for Parameter
  Estimation in High-Mobility OFDM with Fully/Partly Calibrated Massive ULA},''
  \emph{IEEE Transactions on Wireless Communications}, pp. 1--1, 2018.

\bibitem{Monk2016OTFS}
A.~Monk, R.~Hadani, M.~Tsatsanis, and S.~Rakib, ``{OTFS-Orthogonal Time
  Frequency Space},'' 2016.

\bibitem{Ramachandran2018MIMO}
M.~K. Ramachandran and A.~Chockalingam, ``{MIMO-OTFS in High-Doppler Fading
  Channels: Signal Detection and Channel Estimation},'' 2018.

\bibitem{Bellili2014A}
F.~Bellili and S.~Affes, ``{A Low-cost and Robust Maximum Likelihood Doppler
  Spread Estimator},'' in \emph{Global Communications Conference}, 2014, pp.
  4325--4330.

\bibitem{Chizhik2004Slowing}
D.~Chizhik, ``{Slowing the Time-fluctuating MIMO Channel},'' \emph{IEEE
  Transactions on Wireless Communications}, vol.~3, no.~5, pp. 1554--1565,
  2004.

\bibitem{Zhang2012Multiple}
Y.~Zhang, Q.~Yin, P.~Mu, and L.~Bai, ``{Multiple Doppler Shifts Compensation
  and ICI Elimination by Beamforming in High-mobility OFDM systems},'' in
  \emph{International ICST Conference on Communications and NETWORKING in
  China}, 2012, pp. 170--175.

\bibitem{Guo2013Multiple}
W.~Guo, P.~Mu, Q.~Yin, and H.~M. Wang, ``{Multiple Doppler frequency offsets
  compensation technique for high-mobility OFDM uplink},'' in \emph{IEEE
  International Conference on Signal Processing, Communication and Computing},
  2013, pp. 1--5.

\bibitem{Liu2014On}
Z.~Liu, J.~Wu, and P.~Fan, ``{On the Uplink Capacity of High Speed Railway
  Communications with Massive MIMO Systems},'' in \emph{Vehicular Technology
  Conference}, 2014, pp. 1--5.

\bibitem{Chen2017Directivity}
X.~Chen, J.~Lu, T.~Li, P.~Fan, and K.~B. Letaief, ``{Directivity-Beamwidth
  Tradeoff of Massive MIMO Uplink Beamforming for High Speed Train
  Communication},'' \emph{IEEE Access}, vol.~PP, no.~99, pp. 1--1, 2017.

\bibitem{Guo2017High}
W.~Guo, W.~Zhang, P.~Mu, and F.~Gao, ``{High-Mobility OFDM Downlink
  Transmission with Large-Scale Antenna Array},'' \emph{IEEE Transactions on
  Vehicular Technology}, vol.~PP, no.~99, pp. 1--1, 2017.

\bibitem{Guo2018High}
W.~Guo, W.~Zhang, P.~Mu, F.~Gao, and H.~Lin, ``{High-Mobility Wideband Massive
  MIMO Communications: Doppler Compensation, Analysis and Scaling Law},'' 2018.

\bibitem{Guo2018Angle}
W.~Guo, W.~Zhang, P.~Mu, F.~Gao, and B.~Yao, ``{Angle-Domain Doppler
  Pre-Compensation for High-Mobility OFDM Uplink with Massive ULA},'' in
  \emph{GLOBECOM 2017 - 2017 IEEE Global Communications Conference}, 2018, pp.
  1--6.

\bibitem{Wang2009Beam}
J.~Wang, L.~Zhou, C.~W. Pyo, T.~Baykas, and S.~Kato, ``{Beam Codebook Based
  Beamforming Protocol for Multi-Gbps Millimeter-Wave WPAN Systems},''
  \emph{IEEE Journal on Selected Areas in Communications}, vol.~27, no.~8, pp.
  1390--1399, 2009.

\bibitem{Alkhateeb2017Channel}
A.~Alkhateeb, O.~E. Ayach, G.~Leus, and R.~W.~H. Jr, ``{Channel Estimation and
  Hybrid Precoding for Millimeter Wave Cellular Systems},'' \emph{IEEE Journal
  of Selected Topics in Signal Processing}, vol.~8, no.~5, pp. 831--846, 2017.

\bibitem{Gao2016Channel}
Z.~Gao, L.~Dai, H.~Chen, and Z.~Wang, ``{Channel Estimation for Millimeter-Wave
  Massive MIMO with Hybrid Precoding over Frequency-Selective Fading
  Channels},'' \emph{IEEE Communications Letters}, vol.~20, no.~6, pp.
  1259--1262, 2016.

\bibitem{Bajwa2010Compressed}
W.~U. Bajwa, J.~Haupt, A.~M. Sayeed, and R.~Nowak, ``{Compressed Channel
  Sensing: A New Approach to Estimating Sparse Multipath Channels},''
  \emph{Proceedings of the IEEE}, vol.~98, no.~6, pp. 1058--1076, 2010.

\bibitem{Tzikas2008The}
D.~G. Tzikas, C.~L. Likas, and N.~P. Galatsanos, ``{The Variational
  Approximation for Bayesian Inference},'' \emph{IEEE Signal Processing
  Magazine}, vol.~25, no.~6, pp. 131--146, 2008.

\bibitem{Dai2018FDD}
J.~Dai, A.~Liu, and V.~K.~N. Lau, ``Fdd massive mimo channel estimation with
  arbitrary 2d-array geometry,'' \emph{IEEE Transactions on Signal Processing},
  vol.~PP, no.~99, 2018.

\bibitem{Dietrich2000Adaptive}
C.~B. Dietrich, ``{Adaptive Arrays and Diversity Antenna Configurations for
  Handheld Wireless Communication Terminals},'' \emph{Dissertation Abstracts
  International, Volume: 63-08, Section: B, page: 3841.;Chair: Warren L.
  Stutz}, 2000.

\bibitem{Caire_TIT13_JSDM}
A.~Adhikary, J.~Nam, J.-Y. Ahn, and G.~Caire, ``{Joint Spatial Division and
  Multiplexing - The Large-Scale Array Regime},'' vol.~59, no.~10, pp.
  6441--6463, Oct. 2013.

\bibitem{Ji2008Bayesian}
S.~Ji, Y.~Xue, and L.~Carin, ``{Bayesian Compressive Sensing},'' \emph{IEEE
  Transactions on Signal Processing}, vol.~56, no.~6, pp. 2346--2356, 2008.

\bibitem{Wipf2003Bayesian}
D.~P. Wipf and B.~D. Rao, ``{Bayesian Learning for Sparse Signal
  Reconstruction},'' in \emph{IEEE International Conference on Acoustics},
  2003.

\bibitem{Ziniel2013Dynamic}
J.~Ziniel and P.~Schniter, ``{Dynamic Compressive Sensing of Time-Varying
  Signals via Approximate Message Passing},'' \emph{IEEE Transactions on Signal
  Processing}, vol.~61, no.~21, pp. 5270--5284, 2013.

\bibitem{A1977Maximum}
A.~P. Dempster, ``{Maximum Likelihood Estimation from Incomplete Data via the
  EM Algorithm},'' \emph{Journal of the Royal Statistical Society}, vol.~39,
  no.~1, pp. 1--38, 1977.

\bibitem{1Samimi120163}
M.~K. Samimi and T.~S. Rappaport, ``{3-D Millimeter-Wave Statistical Channel
  Model for 5G Wireless System Design},'' \emph{IEEE Transactions on Microwave
  Theory Techniques}, vol.~64, no.~7, pp. 1--19, 2016.

\end{thebibliography}
\end{document}